\documentclass[11pt, a4paper]{article}
\usepackage{jheparxiv}
\usepackage[latin1]{inputenc}
\usepackage{amsmath}
\usepackage{amsfonts}
\usepackage{amssymb}
\usepackage{latexsym}
\usepackage{mathrsfs}
\usepackage{graphicx}
\usepackage{color}
\usepackage{slashed}
\usepackage{twistor}
\usepackage[all]{xy}

\newcommand{\tpsi}{\tilde{\psi}}
\newcommand{\sa}{\mathsf{a}}
\newcommand{\sta}{\tilde{\mathsf{a}}}
\newcommand{\PNC}{\mathbb{P}N}

\renewcommand{\d}{\mathrm{d}}

\subheader{\hfill \begin{tabular}{r} \texttt{IMPERIAL-TP-TA-2017-01}
 \\ \texttt{QMUL-PH-17-02}
 \\ \texttt{CERN-TH-2017-044} \end{tabular}}

\title{Space-time CFTs from the Riemann sphere}

\author[a]{Tim Adamo,}
\author[b,c]{Ricardo Monteiro}
\author[c]{\& Miguel F. Paulos}

\affiliation[a]{Theoretical Physics Group, Blackett Laboratory \\
        Imperial College London, SW7 2AZ, United Kingdom}

\affiliation[b]{Centre for Research in String Theory \\
        Queen Mary University of London, E1 4NS, United Kingdom}

\affiliation[c]{Theoretical Physics Department \\
        CERN, 1211 Geneva 23, Switzerland}

\emailAdd{t.adamo@imperial.ac.uk}
\emailAdd{ricardo.monteiro@qmul.ac.uk}
\emailAdd{miguel.paulos@cern.ch}

\abstract{We consider two-dimensional chiral, first-order conformal field theories governing maps from the Riemann sphere to the projective light cone inside Minkowski space -- the natural setting for describing conformal field theories in two fewer dimensions. These theories have a $\SL(2)$ algebra of local bosonic constraints which can be supplemented by additional fermionic constraints depending on the matter content of the theory. By computing the BRST charge associated with gauge fixing these constraints, we find anomalies which vanish for specific target space dimensions. These critical dimensions coincide precisely with those for which (biadjoint) cubic scalar theory, gauge theory and gravity are classically conformally invariant. Furthermore, the BRST cohomology of each theory contains vertex operators for the full conformal multiplets of single field insertions in each of these space-time CFTs. We give a prescription for the computation of three-point functions, and compare our formalism with the scattering equations approach to on-shell amplitudes.}

\begin{document}
 
\maketitle

\section{Introduction}
\label{Intro}

In recent years we have seen dramatic progress in our understanding of on-shell perturbative observables in a wide array of massless quantum field theories \cite{Elvang:2015rqa}. One striking example are the so-called {\em Cachazo-He-Yuan formulae} (CHY), which express the tree-level, $n$-point scattering amplitudes of a large class of massless QFTs in $d$ dimensions as localized integrals over the moduli space of a $n$-punctured Riemann sphere~\cite{Cachazo:2013hca,Cachazo:2014xea}. The moduli, given by the positions $\{z_i\}$ of the marked points on $\Sigma\cong\CP^1$ up to $\SL(2,\C)$ transformations, are entirely fixed in terms of the kinematic data by a set of constraints known as the \emph{scattering equations}~\cite{Fairlie:1972,Gross:1987ar,Witten:2004cp,Cachazo:2013gna}:
\be\label{Seq}
\sum_{j\neq i}\frac{k_{i}\cdot k_{j}}{z_i -z_j}=0\,,
\ee
where the $\{k_i\}$ are on-shell momenta in $d$ dimensions. Only $n-3$ of these equations are independent, which is precisely the number required to localize all the positions of the marked points on $\Sigma$ since M\"obius invariance trivially fixes three of the $\{z_i\}$.

The CHY formulae give a representation of the tree-level S-matrix for massless QFTs which differs substantially from traditional formulations based on the perturbative expansion of classical space-time actions. Although the formulae can be verified by checking properties such as soft and collinear limits and factorization (e.g. \cite{Dolan:2013isa}), their origin -- and in particular the role of the underlying Riemann sphere -- seems mysterious from the perspective of space-time field theory. This mystery is resolved by \emph{ambitwistor strings}, which are constrained, chiral, first-order 2d CFTs that produce the CHY formulae as sphere correlation functions~\cite{Mason:2013sva}. The precedents for these developments are Witten's seminal proposal of twistor string theory \cite{Witten:2003nn} and the resulting formula for gauge theory scattering amplitudes in four space-time dimensions \cite{Roiban:2004yf}, which are a special case of the new story.

There is by now a small zoo of ambitwistor string theories~\cite{Ohmori:2015sha,Casali:2015vta}, but all are based upon the simple 2d action:
\be\label{ambi1}
S=\frac{1}{2\pi}\int_{\Sigma} \left(P_{\mu}\,\dbar X^{\mu} - \frac{e}{2}\,P^2\right)\,,
\ee
where $X^{\mu}$ are the components of a map from $\Sigma$ to $d$-dimensional (complexified) Minkowski space and $P_{\mu}$ are the conjugate momenta, which have conformal weight $(1,0)$ on $\Sigma$. The Lagrange multiplier field $e$ enforces the constraint $P^2=0$ appropriate to the phase space of massless particles. It is precisely this constraint that generates the scattering equations~\eqref{Seq}. This action has a gauge symmetry generated by the constraint term,
\be\label{ambi2}
\delta X^{\mu}= \alpha\,P^{\mu}\,, \qquad \delta P_{\mu}=0\,, \qquad \delta e=\dbar\alpha\,,
\ee
further reducing the target space from the space of null directions to the space of all null geodesics considered up to scale
,\footnote{$P_{\mu}$ is a 1-form on $\Sigma$, $P_{\mu}=(P_{\mu})_z \d z$, and therefore its component $(P_{\mu})_z$ is only defined up to holomorphic rescaling.} also known as projective ambitwistor space. To quantize \eqref{ambi1} this gauge symmetry must be fixed in addition to holomorphic reparametrization invariance, but this does not lead to any additional anomalies: the only anomaly associated with the ambitwistor string (on a Minkowski background) is the holomorphic conformal anomaly.

All the known ambitwistor strings are modifications of \eqref{ambi1} by the addition of various worldsheet matter systems. Adding two worldsheet current algebras for gauge groups $G$ and $\tilde{G}$ to \eqref{ambi1} leads to a description of biadjoint cubic scalar theory, while a `heterotic' modification of \eqref{ambi1} leads to Yang-Mills theory, and a `type II' modification leads to gravity. These ambitwistor strings do more than just reproduce tree-level amplitude formulae, though: they have led to novel representations for higher-loop field theory integrands in terms of localized expressions both on higher genus Riemann surfaces~\cite{Adamo:2013tsa,Adamo:2015hoa} and on degenerate Riemann spheres~\cite{Geyer:2015bja,Geyer:2015jch,Geyer:2016wjx}.

\medskip

Given the utility of ambitwistor strings for studying on-shell observables for massless QFTs in arbitrary dimension, it seems natural to ask if similar techniques apply to off-shell observables. In this paper, we will be interested in the computation of correlation functions for (classical) \emph{conformal field theories} (CFTs) in space-time of dimension $d$. In particular we will write down actions which are closely related to the ambitwistor string but which, we claim, directly compute correlation functions for various kinds of CFTs. Since CFTs are typically strongly coupled, and have no notion either of particles or `on-shellness', we must temper our ambitions by considering perturbative CFTs. In this case, there is of course a well-known way to obtain correlation functions from the on-shell amplitudes~\cite{fourier}. What our theories accomplish is that they appear to compute correlators directly without passing through the amplitudes. While `on-shell-like' methods have been introduced for computing correlators in $\mathcal N=4$ SYM theory (cf. \cite{Adamo:2011dq,Adamo:2011cd,Chicherin:2014uca,Chicherin:2016fbj,Koster:2016ebi,Koster:2016loo,Chicherin:2016qsf,He:2016dol,Brandhuber:2016xue,He:2016jdg,Koster:2016fna,Eden:2017fow}), our models can in principle compute correlators in theories such as $\phi^3$ in $d=6$. In this paper however, we mostly focus on the general structure of the theories, most notably their BRST quantization. 

The complexified conformal group SO$(d+2,\C)$ acts non-trivially on $d$-dimensional Minkowski space, but it is well-known that this action is linearized on the \emph{projective null cone} in $D:=d+2$ dimensions, where it is simply the action of the Lorentz group. The models we consider are chiral 2d CFTs governing holomorphic maps from the Riemann sphere to the cotangent space of the projective null cone, which is the natural phase space for describing CFTs in $d$ dimensions. Unlike ambitwistor strings, these models have a non-abelian triplet of constraints which can be grouped together to form a non-dynamical $\SL(2)$ gauge field on the Riemann sphere. Our construction can be thought of as a chiral complexification of the Marnelius particle model~\cite{Marnelius1979} or the `two-time physics' of Bars~\cite{Bars:1997bz,Bars:1997xb,Bars:1998ph,Bars:1998pc,Bars:2000qm}.\footnote{Our work also invites interesting comparisons with Green's `worldsheets for worldsheets'~\cite{Green:1987cw}, especially in the case of gravity.} As emphasized there, different gauges can provide alternative descriptions of equivalent physics. In particular, one special gauge reduces our model to the ordinary ambitwistor string, but demanding that the $\SL(2)$ gauge symmetry is not anomalous fixes the target space dimension to a definite value. Hence in this context, the gauge invariance simply implies the equivalence between on-shell amplitudes and correlation functions for perturbative theories.

We study three versions of our theories, which are referred to as the \emph{bosonic}, \emph{heterotic} and \emph{type II} models. In the heterotic and type II models, the non-dynamical $\SL(2)$ gauge field is supplemented by additional fermionic constraints so that the target space becomes a supersymmetric version of the projective null cone. The quantum properties of these 2d models encode the classical conformal invariance of theories on the target space. In particular, we find that each model encodes information about a specific (classical) space-time CFT: the bosonic model leads to $d=6$ biadjoint cubic scalar theory; the heterotic model leads to $d=4$ gauge theory; and the type II model leads to $d=2$ gravity. These connections are made by computing anomalies in the 2d models which produce the critical dimensions required for classical conformal invariance in the space-time theories; by investigating the vertex operator spectra of the models; and by considering three-point functions.

After a brief review of the projective null cone and its associated phase space in section \ref{PNC}, we introduce the three models and study their classical symmetries in section \ref{CMod}. We then gauge fix these symmetries in section \ref{QMod} and show that the gauge anomalies are killed in certain critical dimensions of the target space for each model. The BRST cohomology is also seen to contain vertex operators which encode all single field insertions of the relevant space-time CFTs. Section \ref{3pt} explores a prescription for the computation of three-point functions in the space-time CFT from the three-point correlators of these models on the Riemann sphere. We conclude with a discussion of the many open questions and unresolved issues raised throughout this paper in section \ref{Disc}.


\section{The Projective Null Cone}
\label{PNC}

Consider complexified $d$-dimensional Euclidean space; this is simply $\C^d$ with the flat, holomorphic Euclidean metric. The conformal group, SO$(d+1,1)$ complexified to SO$(d+2,\C)$, acts in a non-trivial way on this space. The study of classical conformal field theories (CFTs) is facilitated by going to a space which linearizes the action of the conformal group. This is achieved by considering the action of the conformal group and formulating CFTs on the \emph{projective null cone} in two higher dimensions (cf., \cite{Dirac:1936fq,Mack:1969rr,Ferrara:1973eg}). We will briefly review this construction here, as well as the associated phase space on which our models will live.

Denote by $D:=d+2$ the dimension of the `embedding space' $\C^{D}$, endowed with the flat metric on coordinates $X^{\mu}=(X^{+},X^{-},x^{a})$
\begin{equation*}
 \d s^2 = \eta_{\mu\nu}\, \d X^{\mu}\,\d X^{\nu}=-\d X^{+}\, \d X^{-} + \d x^{2}\,,
\end{equation*}
where $a=0,\ldots,d-1$. The null cone defined by $X^2=0$ is a SO$(d+1,1)$-invariant subspace of the embedding space $\C^{D}$, and the projective null cone, $\PNC$ is obtained by quotienting by scale:
\be\label{NC1}
\PNC=\left\{X\in \C^{D}| X^2=0\right\}/\Upsilon\,, \qquad \Upsilon:=X\cdot\frac{\partial}{\partial X}\,.
\ee
The action of the Lorentz group SO$(d+1,1)$ descends to $\PNC$ since the action of the Euler vector field $\Upsilon$ respects Lorentz rotations. In other words, $\PNC$ is a $d$-dimensional space with a natural action of SO$(d+1,1)$.

To see that this action is equivalent to the action of the conformal group in $d$ dimensions, we can consider a particular coordinate patch of the projective space $\PNC$.\footnote{In the embedding space literature, this is often referred to as `picking a section.'} We label this patch by a choice of `infinity vector', $I^\mu$, such that $I\cdot X\neq 0$. For instance, consider the coordinate patch where $X^{+}\neq 0$; since the $X^{\mu}$ are homogeneous coordinates on $\PNC$ (being defined only up to scale) we can normalize by $X^{+}$ to consider coordinates $X^{\mu}=(1,X^{-},x^a)$. The $X^{2}=0$ condition then enforces $X^{-}=x^2$ on this coordinate patch, so the only degrees of freedom in $X^{\mu}=(1,x^2, x^a)$ are those of the `physical' space $\C^d$. A general Lorentz transformation $\Lambda^{\mu}_{\;\nu}$ in $D$ dimensions induces a transformation of $x^a=X^\mu/X^+$ which is precisely a conformal transformation. That is, the linear action of the Lorentz group on the embedding space descends to the non-linear action of the conformal group on the physical space.

Conformal primaries in $d$ dimensions can also be written simply as tensors on the projective null cone~\cite{Dirac:1936fq,Ferrara:1973eg,Costa:2011mg}. Indeed, a spin $s$ conformal primary operator of conformal dimension $\Delta$ on $\C^d$ is given in terms of a tensor field $T_{\mu_{1}\cdots\mu_{s}}(X)$ defined on $\PNC$ which is totally symmetric, traceless and obeys
\be\label{NCf}
X^{\mu_1}\,T_{\mu_1\cdots\mu_s}(X)=0\,, \qquad T_{\mu_1\cdots\mu_s}(\lambda X)= \lambda^{-\Delta}\,T_{\mu_1\cdots\mu_s}( X)\,.
\ee
Furthermore, any $D$-dimensional tensor is only defined up to shifts by lower rank tensors times $X^\mu$, since this will not affect the $d$-dimensional result. For instance, we have the identifications between $D$ and $d$-dimensional fields,
\be\label{phi}
\phi(x)=(I\cdot X)^{\Delta} \Phi(X) \bigg |_{X^2=0}\,, \qquad V_a(x)= (I\cdot X)^{\Delta-1}\frac{\partial X^\mu}{\partial x^{a}} V_\mu(X) \bigg |_{X^2=0}\,.
\ee
Furthermore, any $d$-dimensional quantity, such as derivatives, can be obtained from embedding space by the use of a projector. This is further explained in appendix \ref{appendix}.

Since $\PNC$ is a projective space and the conformal primaries are just tensors of fixed homogeneity, conformal integrals over $\PNC$ are constrained (up to an overall factor) by homogeneity, leading to myriad applications of the projective null cone, and the embedding space more generally, to the study of CFTs in $d>2$ (see for instance \cite{Weinberg2010b,Weinberg2012b,Siegel2012,Paulos:2011ie,Goldberger:2011yp,SimmonsDuffin:2012uy,Paulos:2012qa}).

\medskip

This simplicity makes the projective null cone the natural space to study theories with conformal invariance. We will be particularly interested in the \emph{phase space} associated with $\PNC$, which will become the target space of our 2d models. Generally speaking, the phase space will be a subset of the cotangent bundle, charted with canonical coordinates $(X,P)$. For example, in the case relevant for ambitwistor strings -- namely, massless particles -- the natural phase space is:
\be\label{ND}
T^{*}_{N}:=\left\{\left.(x,p)\in T^{*}\C^{d}\:\right|\: p^2=0\right\}\,.
\ee
For the study of CFTs in $\C^d$, the natural phase space is the one associated with the projective null cone in $\C^{D}$. This is given by:
\be\label{NC3}
T^{*}\PNC:=\left\{\left.(X,P)\in T^{*}\C^{D}\:\right|\: X^{2}=X\cdot P=P^{2}=0\right\}/\left\{X\cdot\frac{\partial}{\partial X},\:P\cdot\frac{\partial}{\partial P}\right\}\,.
\ee
The constraints $X\cdot P=0$ and $P^2=0$ ensure that motion in the phase space is on and tangent to the projective null cone, rather than moving off the quadric into $\C^{D}$. We also quotient by the scales of $X^{\mu}$ and $P_{\nu}$ so that $T^{*}\PNC$ is projective. Note that the phase space of massless particles in $d$ dimensions has the same dimensionality as the phase space of the projective null cone in $D$ dimensions,
\begin{equation*}
 \mathrm{dim}(T^{*}_{N})=2d-1=2D-5= \mathrm{dim}(T^{*}\PNC)\,,
\end{equation*}
when $D=d+2$.

We can be very explicit in showing that $T^{*}_{N}\cong T^{*}\PNC$. Since $X,P$ appear on the same footing, we first solve the $X^2=P^2=0$ constraints and fix the scalings to write $X^\mu=(1,x^2,x^a)$, $P^\mu=(1,y^2,y^a)$. We then have $P\cdot X=-(x-y)^2/2=0$, so $x^a,y^a$ are null separated.  Since this is the case, we can keep $x^a$ and trade $y^a$ for the null vector $p^a=y^a-x^a$, which shows the equivalence of the phase spaces. Equivalently, in embedding space language, we can notice that given a solution $X,P$ of the constraints, we can define new solutions by performing shifts $X$ by $P$ and vice versa. Hence we can define $P\to P-X$ which gives $P^\mu=(0,2p\cdot x, p^a)$, with null $p^a=y^a-x^a$.

In fact, we will be interested in a further reduction of these spaces. In the case of the massless particle, the space of null geodesics is a quotient of $T^{*}_{N}$ by the action of $p\cdot \partial_x$, so that points along the same geodesic are identified; this is \emph{ambitwistor space}, with complex dimension $2d-2$. The effective target space of ambitwistor strings requires a further quotient by scale of $p^\mu$ (i.e. by the action of $p\cdot\partial_p$); this is \emph{projective ambitwistor space}, with complex dimension $2d-3$. For $T^{*}\PNC$, the equivalent reduction is achieved by quotienting by the freedom to shift $X$ by $P$ and vice-versa; this enlarges $\{X\cdot\partial_X,\,P\cdot\partial_P\}$ in the quotient \eqref{NC3} to $\{X\cdot\partial_X,P\cdot\partial_P,P\cdot\partial_X,X\cdot\partial_P\}$.

We will see later how the equivalence of  $T^{*}_{N}$ and $T^{*}\PNC$, and of their respective reductions, provides the connection between the models to be proposed here and ambitwistor strings.

There are straightforward generalizations of the projective null cone which allow for supersymmetry. If the coordinates $X^{\mu}$ on $\C^{D}$ are supplemented with $(\psi^{\mu}_{1},\ldots,\psi^{\mu}_{r})$, then the null cone includes super-null directions: $X\cdot\psi_{1}=\cdots=X\cdot\psi_{r}=0$. The associated phase space is then given by constraining the extended cotangent bundle:
\begin{equation*}
 T^{*}_{(r)}\C^{D}\cong T^{*}\C^{D}\bigoplus_{i=1}^{r} \Pi T\C^{D}\,,
\end{equation*}
where $\Pi$ is the parity reversing functor. For each new fermionic direction, $\psi^{\mu}_{i}$, there are two additional fermionic constraints: one on the super-null cone ($\psi_i \cdot X=0$) and one on the fibres ($\psi_i \cdot P=0$). In addition, there are new bosonic constraints which enforce $\psi_{i}\cdot\psi_{j}=0$ for all $i\neq j$.


\section{Theories on the Riemann sphere -- classical aspects}
\label{CMod}

To write a chiral, first-order 2d CFT governing maps from the Riemann sphere to the phase space of the projective null cone, one simply adds Lagrange multiplier terms to the action for a free chiral boson which enforce the appropriate constraints. In this section, we introduce three such theories which will be referred to as the \emph{bosonic}, \emph{heterotic}, and \emph{type II} models. These names reflect the constraint algebras of the 2d CFTs and their matter content, although it must be emphasized that these models are strictly chiral: they contain only left-moving degrees of freedom on the Riemann sphere.


\subsection{The bosonic model}

Let $\Sigma\cong\CP^{1}$ and $X^{\mu}:\Sigma\rightarrow \C^{D}$ be a map from the Riemann sphere to (complexified) $D$-dimensional Minkowski space. The bosonic model is defined by the 2d action:
\be\label{bm1}
S=\frac{1}{2\pi}\int_{\Sigma} \left( P_{\mu}\,\dbar X^{\mu} -\frac{e^{(1)}}{2}\,P^2 -\frac{e^{(2)}}{2}\,X^{2} -e^{(3)}\,P\cdot X\right) + S_{\mathfrak{g}} + S_{\tilde{\mathfrak{g}}}\,,
\ee
where $P_{\mu}$ is the holomorphic conjugate momentum of $X^{\mu}$. Note that while $X^\mu$ is a set of functions, $P_{\mu}$ carries conformal weight
 $(1,0)$, and is really a space-time covector taking values in one-forms on $\Sigma$. Equivalently, $P_\mu$ takes values in $\Omega^{0}(\Sigma, K_{\Sigma}\otimes\C^{D})$; in local coordinates $z$ on $\Sigma$, $P_{\mu}= (P_{\mu})_{z}\d z$. The action $S_{\mathfrak{g}}$ describes a \emph{holomorphic} worldsheet current algebra (consisting of free fermions, a WZW model or any other construction) for the Lie algebra $\mathfrak{g}$. $S_{\tilde{\mathfrak{g}}}$ is a second holomorphic worldsheet current algebra, where $\mathfrak{g}$, $\tilde{\mathfrak{g}}$ can be distinct. The $(P,X)$ part of the action can be viewed as a holomorphic complexification of the worldline action introduced by Marnelius \cite{Marnelius1979} and developed by Bars in the context of `two-time physics'~\cite{Bars:1997bz,Bars:1997xb,Bars:1998ph,Bars:1998pc,Bars:2000qm}. Equivalently, it can be thought of as a further-constrained version of the $D$-dimensional ambitwistor string~\cite{Mason:2013sva}.

The fields $\{e^{(1)},e^{(2)}, e^{(3)}\}$ serve as Lagrange multipliers enforcing the constraints $P^{2}=X^{2}=X\cdot P=0$. Since $P_{\mu}$ is a holomorphic 1-form, it follows that the Lagrange multipliers must carry non-trivial conformal weights in order for the action \eqref{bm1} to make sense. In particular, the conformal weights (or bundle values) of the Lagrange multiplier fields are given by:
\begin{center}
 \begin{tabular}{||c c c||} 
 \hline
 Lagrange multiplier & Conformal weights & Bundle \\ [0.5ex] 
 \hline\hline
 $e^{(1)}$ & $(-1,1)$ & $\Omega^{0,1}(\Sigma, T_{\Sigma})$ \\ 
 \hline
 $e^{(2)}$ & $(1,1)$ & $\Omega^{0,1}(\Sigma, K_{\Sigma})$ \\
 \hline
 $e^{(3)}$ & $(0,1)$ & $\Omega^{0,1}(\Sigma)$ \\ [1ex] 
 \hline
\end{tabular}
\end{center}
In the action \eqref{bm1}, we have made a canonical choice of complex structure on $\Sigma$. The theory is in fact a (chiral, first-order) 2d CFT.

\medskip

The fields $X^{\mu}$ and $P_{\mu}$ have classical Poisson brackets
\be\label{bpb}
\left\{X^{\mu}\,,\;P_{\nu}\right\}=\delta^{\mu}_{\nu}\,, \qquad \left\{X^{\mu}\,,\;X^{\nu}\right\}=0=\left\{P_{\mu}\,,\;P_{\nu}\right\}\,,
\ee
under which the algebra of constraints is closed,
\be\label{constrpb}
\left\{X^2\,,\;P^2\right\}=4\,X\cdot P\,, \qquad 
\left\{X\cdot P\,,\;X^2\right\}=-2\,X^2\,, \qquad 
\left\{X\cdot P\,,\;P^2\right\}=2\,P^2\,.
\ee
Associated to the constraints, there are additional symmetries of the action, beyond holomorphic reparametrization invariance. These symmetries lead to a further reduction in the target space of the theory, on top of the constraints. A straightforward calculation demonstrates that~\eqref{bm1} is invariant under the linearised transformations:
\begin{subequations}\label{btrans1}
\begin{eqnarray}
\delta X^{\mu} & = & \alpha^{(1)}\,P^{\mu} +\alpha^{(3)}\,X^{\mu}\,, \\ 
\delta P_{\mu} & = & -\alpha^{(2)}\,X_{\mu} - \alpha^{(3)}\,P_{\mu}\,,
\end{eqnarray}
\end{subequations}
\begin{subequations}\label{btrans2}
\begin{eqnarray}
 \delta e^{(1)} & = & \dbar \alpha^{(1)}+2\,e^{(1)}\alpha^{(3)} - 2\,e^{(3)}\alpha^{(1)}\,, \\
 \delta e^{(2)} & = & \dbar \alpha^{(2)}-2\,e^{(2)}\alpha^{(3)} + 2\,e^{(3)}\alpha^{(2)}\,, \\
 \delta e^{(3)} & = & \dbar \alpha^{(3)} + e^{(1)}\alpha^{(2)} - e^{(2)}\alpha^{(1)}\,.
\end{eqnarray}
\end{subequations}
Here, the gauge parameters $\{\alpha^{(1)},\alpha^{(2)},\alpha^{(3)}\}$ have conformal weights $(-1,0)$, $(1,0)$, and $(0,0)$ respectively (these can be read off by requiring that the transformations \eqref{btrans1}--\eqref{btrans2} have definite conformal weight).

The transformations \eqref{btrans1} are the worldsheet realization of the reduction via quotients discussed in \eqref{NC3} and below, leading to an effective target space with complex dimension $2D-7$. In particular, the transformations with parameters $\alpha^{(1)}$, $\alpha^{(2)}$ and $\alpha^{(3)}$ are associated, respectively, with the actions of $P\cdot\partial_X$, $-X\cdot\partial_P$ and $X\cdot\partial_X-P\cdot\partial_P$. In the latter case, the extension to the independent actions of $X\cdot\partial_X$ and $P\cdot\partial_P$ is provided by the fact that the component $(P_\mu)_z$ of the 1-form $P_\mu$ on $\Sigma$ is ab initio only defined up to holomorphic rescalings\footnote{Notice that the generators $V^{(a)}$ represent the Hamiltonian vector fields associated to the constraints $h^{(a)}$, that is, $\{h^{(a)},\cdot\}=-V^{(a)}$. As standard, in the 2d CFT language, this action is realised via $\oint h^{(a)}$, and the restriction to the classical worldsheet theory is the restriction to single contractions in the operator product expansion. In the quantum theory, the action $\oint Q$ of the BRST operator, to be defined later for our models, is required.}.

An important feature of these additional symmetries of the action is that they are \emph{non-abelian} in nature: viewing $e^{(a)}$ (for $a=1,2,3$) as a gauge field, the transformation $\delta e^{(a)}$ involves all three of the gauge parameters, not just $\alpha^{(a)}$. This non-abelian structure can be made manifest by noting that the triplet of constraints form a representation of $\mathfrak{sl}(2)$ under the action of the Poisson bracket \eqref{bpb}. More generally, the action \eqref{bm1} can be written explicitly in terms of a non-dynamical $\SL(2)$ gauge field on $\Sigma$. Let $\alpha,\beta,\ldots=1,2$ denote two-component $\SL(2)$ spinor indices which are raised and lowered with the two-dimensional Levi-Civita symbols $\epsilon_{\alpha\beta}$, $\epsilon^{\alpha\beta}$,  and define:
\be\label{bm2}
Z^{\alpha}_{\mu}=\left(X_{\mu}\,, \; P_{\mu}\right)\,, \qquad A_{\alpha\beta}=\left(
\begin{array}{cc}
 e^{(2)} & e^{(3)} \\
 e^{(3)} & e^{(1)}
\end{array}\right)\,.
\ee
Note that different components of $Z^{\alpha}_{\mu}$ and $A_{\alpha\beta}$ have differing conformal weights on the Riemann sphere, so these objects are not tensorial on $\Sigma$. However, they do enable a compact re-writing of the action \eqref{bm1} as:
\be\label{bm3}
S=-\frac{1}{4\pi}\int_{\Sigma} \eta^{\mu\nu} Z^{\alpha}_{\mu}\left(\epsilon_{\alpha\beta}\,\dbar + A_{\alpha\beta}\right) Z^{\beta}_{\nu}\,,
\ee
where $\eta^{\mu\nu}$ is the flat metric in $D$ space-time dimensions. 

Written in this way, the action is invariant under $\SL(2)$ gauge transformations, $M^{\alpha}{}_{\beta}$:
\be\label{nab1}
Z^{\alpha}_{\mu}\rightarrow M^{\alpha}{}_{\beta}\,Z^{\beta}_{\mu}\,, \qquad A_{\alpha\beta}\rightarrow M_{\alpha}{}^{\gamma}\,A_{\gamma\delta}\,M_{\beta}{}^{\delta} - M_{\beta}{}^{\gamma}\,\dbar M_{\alpha\gamma}\,.
\ee
These non-abelian transformations can be seen as the exponentiation of the infinitesimal symmetries \eqref{btrans1}--\eqref{btrans2} by parametrizing
\begin{equation*}
 M^{\alpha}{}_{\beta}=\exp\left(m^{\alpha}{}_{\beta}\right)\,, \qquad m^{\alpha}{}_{\beta}=\left(\begin{array}{cc}
                                                                                                            \alpha^{(3)} & \alpha^{(1)} \\
                                                                                                            -\alpha^{(2)} & -\alpha^{(3)}
                                                                                                           \end{array}\right)\,,
\end{equation*}
and then linearizing the full transformations \eqref{nab1}. Hence, the bosonic model is equivalent to a system of chiral bosons in 2d coupled to a non-dynamical $\SL(2)$ gauge field.

The non-abelian gauge symmetries present a large amount of gauge redundancy -- at least, by comparison with the ambitwistor string, whose symmetries are abelian in nature~\cite{Mason:2013sva,Adamo:2013tsa,Ohmori:2015sha}. The gauge-fixing and quantization associated with these issues will be addressed in Section~\ref{QMod}.


\subsection{The heterotic model}

As the nomenclature suggests, the heterotic model is a modification of the bosonic model. Replacing one copy of the worldsheet current algebra by a system of fermionic spinors, the 2d action becomes:
\begin{multline}\label{het1}
S=\frac{1}{2\pi}\int_{\Sigma} P_{\mu}\,\dbar X^{\mu} -\frac{1}{2}\psi_{\mu}\,\dbar\psi^{\mu} + S_{\mathfrak{g}} \\
-\frac{e^{(1)}}{2}\,P^2 -\frac{e^{(2)}}{2}\,X^{2} -e^{(3)}\,P\cdot X +\chi^{(1)}\,\psi\cdot P +\chi^{(2)}\,\psi\cdot X \,,
\end{multline}
where $X^{\mu}$, $P_{\mu}$, $\{e^{(1)}, e^{(2)}, e^{(3)}\}$, and $S_{\mathfrak{g}}$ are as before. The real fermions $\psi^{\mu}$ have conformal weight $(\frac{1}{2},0)$, or equivalently take values in $\Omega^{0}(\Sigma, K^{1/2}_{\Sigma}\otimes\C^{D})$. The system of constraints is now adapted to the supersymmetrized phase space, with the bosonic constraints augmented by two fermionic constraints $\psi\cdot P=0=\psi\cdot X$, imposed by the Lagrange multipliers $\chi^{(1)},\chi^{(2)}$.

These Lagrange multipliers have fermionic statistics and carry conformal weight on the Riemann sphere: $\chi^{(1)}$ has conformal weight $(-\frac{1}{2},1)$ and $\chi^{(2)}$ has conformal weight $(\frac{1}{2},1)$. Just as in the bosonic model, the various constraints appearing in the second line of \eqref{het1} generate additional symmetries of the action beyond holomorphic reparametrization invariance. The bosonic symmetries of \eqref{btrans1}--\eqref{btrans2}, and more generally \eqref{nab1}, are augmented by the transformations
\be\label{htrans1}
\delta\chi^{(1)}=\chi^{(1)}\,\alpha^{(3)} - \chi^{(2)}\,\alpha^{(1)}\,, \qquad \delta\chi^{(2)}=\chi^{(1)}\,\alpha^{(2)}-\chi^{(2)}\,\alpha^{(3)}\,,
\ee
but otherwise unchanged. In addition to these bosonic symmetries, \eqref{het1} possesses fermionic symmetries:
\begin{subequations}\label{htrans2}
\begin{eqnarray}
\delta X^{\mu} & = & \varepsilon^{(1)}\,\psi^{\mu} \,, \\ 
\delta P_{\mu} & = & -\varepsilon^{(2)}\,\psi_{\mu} \,, \\
\delta \psi^{\mu} & = & \varepsilon^{(1)}\,P^{\mu} + \varepsilon^{(2)}\,X^{\mu} \,,
\end{eqnarray}
\end{subequations}
\begin{subequations}\label{htrans3}
\begin{eqnarray}
 \delta e^{(1)} & = & \chi^{(1)}\,\varepsilon^{(1)}\,, \\
 \delta e^{(2)} & = & \chi^{(2)}\,\varepsilon^{(2)}\,, \\
 \delta e^{(3)} & = & \chi^{(1)}\,\varepsilon^{(2)} +\chi^{(2)}\,\varepsilon^{(1)}\,, \\
 \delta \chi^{(1)} & = & - \dbar \varepsilon^{(1)} - e^{(1)}\,\varepsilon^{(2)} + e^{(3)}\,\varepsilon^{(1)}\,, \\
 \delta \chi^{(2)} & = & - \dbar \varepsilon^{(2)} + e^{(2)}\,\varepsilon^{(1)} - e^{(3)}\,\varepsilon^{(2)}\,.
\end{eqnarray}
\end{subequations}
The new gauge parameters $\{\varepsilon^{(1)},\varepsilon^{(2)}\}$ have fermionic statistics and conformal weights $(-\frac{1}{2},0)$ and $(\frac{1}{2},0)$, respectively. 

Due to their fermionic nature, these new symmetries cannot be exponentiated: they are fully general in their infinitesimal form. The non-dynamical gauge group of the heterotic model is the Lie supergroup $\mathrm{OSp}(2|1)$, which has super-dimension $(6|2)$ and bosonic body $\mathrm{Sp}(2)\cong\SL(2)$~(cf. \cite{deWitt:1992}). 


\subsection{The type II model}

The final variant, as the nomenclature suggest, replaces the remaining worldsheet current algebra of the heterotic model with a second system of fermionic spinors on the Riemann sphere. The 2d action for the type II model reads:
\begin{multline}\label{II1}
 S=\frac{1}{2\pi}\int_{\Sigma} P_{\mu}\,\dbar X^{\mu} -\frac{1}{2}\psi_{\mu}\,\dbar\psi^{\mu} -\frac{1}{2}\tpsi_{\mu}\,\dbar\tpsi^{\mu} -\frac{e^{(1)}}{2}\,P^2 -\frac{e^{(2)}}{2}\,X^{2} -e^{(3)}\,P\cdot X \\
 +\chi^{(1)}\,\psi\cdot P +\chi^{(2)}\,\psi\cdot X +\tilde{\chi}^{(1)}\,\tpsi\cdot P +\tilde{\chi}^{(2)}\,\tpsi\cdot X +\xi\,\psi\cdot\tpsi\,.
\end{multline}
In contrast to type II string theory, the new set of fermions $\tpsi^{\mu}$ in this model have the \emph{same} chirality as the $\psi^{\mu}$; that is, they have conformal weight $(\frac{1}{2},0)$, or take values in $\Omega^{0}(\Sigma, K^{1/2}_{\Sigma}\otimes \C^{D})$. The set of constraints is augmented so as to include the `tilded' analogues of the fermionic constraints introduced in the heterotic model, namely $\tpsi\cdot P = 0 =\tpsi\cdot X$. The associated Lagrange multipliers, $\tilde{\chi}^{(1)}$ and $\tilde{\chi}^{(2)}$, have the same statistics and conformal weights as their un-tilded cousins.

To obtain a closed constraint algebra, an additional bosonic constraint $\psi\cdot\tpsi=0$ is enforced by a bosonic Lagrange multiplier $\xi$ with conformal weight $(0,1)$. This results in an extended set of bosonic symmetries for \eqref{II1}, given by \eqref{btrans1}--\eqref{btrans2} and
\begin{subequations}\label{IItrans1}
\begin{eqnarray}
 \delta \psi^{\mu} & = & -\rho\,\tpsi^{\mu}\,, \\
 \delta \tpsi^{\mu} & = & \rho\,\psi^{\mu}\,, \\
 \delta \chi^{(1)} & = & \chi^{(1)}\,\alpha^{(3)} - \chi^{(2)}\,\alpha^{(1)} - \tilde{\chi}^{(1)}\,\rho\,, \\
 \delta \chi^{(2)} & = & \chi^{(1)}\,\alpha^{(2)} - \chi^{(2)}\,\alpha^{(3)} - \tilde{\chi}^{(2)}\,\rho\,, \\
 \delta \tilde{\chi}^{(1)} & = & \tilde{\chi}^{(1)}\,\alpha^{(3)} - \tilde{\chi}^{(2)}\,\alpha^{(1)} + \chi^{(1)}\,\rho\,, \\
 \delta \tilde{\chi}^{(2)} & = & \tilde{\chi}^{(1)}\,\alpha^{(2)} - \tilde{\chi}^{(2)}\,\alpha^{(3)} + \chi^{(2)}\,\rho\,, \\
 \delta \xi & = & - \dbar\rho\,,
\end{eqnarray}
\end{subequations}
where $\rho$ is the gauge parameter associated with the transformations generated by the $\psi\cdot\tpsi=0$ constraint, having conformal weight $(0,0)$. The fermionic symmetries of \eqref{II1} are:
\begin{subequations}\label{IItrans2}
\begin{eqnarray}
\delta X^{\mu} & = & \varepsilon^{(1)}\,\psi^{\mu} + \tilde{\varepsilon}^{(1)}\,\tpsi^{\mu} \,, \\ 
\delta P_{\mu} & = & -\varepsilon^{(2)}\,\psi_{\mu} - \tilde{\varepsilon}^{(2)}\,\tpsi_{\mu}\,, \\
\delta \psi^{\mu} & = & \varepsilon^{(1)}\,P^{\mu} + \varepsilon^{(2)}\,X^{\mu} \,, \\
\delta \tpsi^{\mu} & = & \tilde{\varepsilon}^{(1)}\,P^{\mu} + \tilde{\varepsilon}^{(2)}\,X^{\mu}\,.
\end{eqnarray}
\end{subequations}
\begin{subequations}\label{IItrans3}
\begin{eqnarray}
 \delta e^{(1)} & = & \chi^{(1)}\,\varepsilon^{(1)} + \tilde{\chi}^{(1)}\,\tilde{\varepsilon}^{(1)}\,, \\
 \delta e^{(2)} & = & \chi^{(2)}\,\varepsilon^{(2)} + \tilde{\chi}^{(2)}\,\tilde{\varepsilon}^{(2)}\,, \\
 \delta e^{(3)} & = & \chi^{(1)}\,\varepsilon^{(2)} +\chi^{(2)}\,\varepsilon^{(1)} +\tilde{\chi}^{(1)}\,\tilde{\varepsilon}^{(2)}+\tilde{\chi}^{(2)}\,\tilde{\varepsilon}^{(1)}\,, \\
 \delta \chi^{(1)} & = & - \dbar \varepsilon^{(1)} - e^{(1)}\,\varepsilon^{(2)} + e^{(3)}\,\varepsilon^{(1)} + \xi\,\tilde{\varepsilon}^{(1)}\,, \\
 \delta \chi^{(2)} & = & - \dbar \varepsilon^{(2)} + e^{(2)}\,\varepsilon^{(1)} - e^{(3)}\,\varepsilon^{(2)} + \xi\,\tilde{\varepsilon}^{(2)}\,, \\
 \delta \tilde{\chi}^{(1)} & = & - \dbar \tilde{\varepsilon}^{(1)} - e^{(1)}\,\tilde{\varepsilon}^{(2)} + e^{(3)}\,\tilde{\varepsilon}^{(1)} + \xi\,\varepsilon^{(1)}\,, \\
 \delta \tilde{\chi}^{(2)} & = & - \dbar \tilde{\varepsilon}^{(2)} + e^{(2)}\,\tilde{\varepsilon}^{(1)} - e^{(3)}\,\tilde{\varepsilon}^{(2)} + \xi\,\varepsilon^{(2)}\,, \\
 \delta \xi & = & \tilde{\chi}^{(1)}\,\varepsilon^{(2)} - \tilde{\chi}^{(2)}\,\varepsilon^{(1)} + \chi^{(2)}\,\tilde{\varepsilon}^{(1)} - \chi^{(1)}\,\tilde{\varepsilon}^{(2)}\,.
\end{eqnarray}
\end{subequations}
The fermionic gauge parameters $\tilde{\varepsilon}^{(1)}$ and $\tilde{\varepsilon}^{(2)}$ have the same conformal weights as their un-tilded counterparts.

\medskip

The gauge group of the type II model is $\SL(2|1)$, which has bosonic body $\SL(2)\times\C^*$ and super-dimension $(4|4)$. Since the type II constraint algebra seems to be the most general one associated with the projective null cone in $\C^{D}$, we summarize it here:
\begin{center}
 \begin{tabular}{||c c c c||} 
 \hline
 Lagrange multiplier & Conformal weights & Statistics & Associated Constraint \\ [0.5ex] 
 \hline\hline
 $e^{(1)}$ & $(-1,1)$ & Bosonic & $P^2$ \\ 
 \hline
 $e^{(2)}$ & $(1,1)$ & Bosonic & $X^2$ \\
 \hline
 $e^{(3)}$ & $(0,1)$ & Bosonic & $X\cdot P$ \\
 \hline
 $\chi^{(1)}, \tilde{\chi}^{(1)}$ & $(-\frac{1}{2},1)$ & Fermionic & $\psi\cdot P, \tpsi\cdot P$ \\
 \hline
 $\chi^{(2)}, \tilde{\chi}^{(2)}$ & $(\frac{1}{2},1)$ & Fermionic & $\psi\cdot X, \tpsi\cdot X$ \\
 \hline
 $\xi$ & $(0,1)$ & Bosonic & $\psi\cdot\tpsi$ \\ [1ex]
 \hline
\end{tabular}
\end{center}
The fermionic constraints can be viewed as simply taking two copies of the fermionic constraints from the heterotic model. These two copies are then tied together by the fourth bosonic constraint $\psi\cdot\tilde{\psi}$, which can be thought of as the generator of an $\mathrm{O}(2)$ symmetry rotating tilded fields into un-tilded fields, and vice versa.


\section{Quantization, Critical Dimensions \& Vertex Operators}
\label{QMod}

Each of the three models described in Section~\ref{CMod} has a non-abelian set of gauge redundancies, building on the $\SL(2)$ set of bosonic constraints which restrict the target space to the projective light cone in $D$ dimensions. To quantize the models these gauge redundancies must be fixed; a natural choice of gauge is one in which the two-dimensional actions become free. This corresponds to gauge-fixing all the Lagrange multiplier fields to zero. The resulting BRST operators allow us to compute all anomalies associated with the gauge fixing exactly, as well as to make some statements about the vertex operator spectrum of each model.
In addition, all three models are holomorphic conformal field theories in two dimensions, so there is also the option to gauge fix chiral gravity on the Riemann surface. Doing so opens the door to interpreting these models as string theories in their own right, although there are difficulties with this as we will discuss.


\subsection{BRST quantization}

Consider the bosonic model \eqref{bm1} with its set of non-abelian gauge symmetries: \eqref{btrans1}--\eqref{btrans2} in the infinitesimal case or \eqref{nab1} in the finite case. These symmetries can be used to gauge-fix the values of the Lagrange multiplier fields $e^{(1)}$, $e^{(2)}$, and $e^{(3)}$ to zero. Equivalently, in the manifest $\SL(2)$ formulation \eqref{bm3} this is a gauge in which the non-dynamical gauge field vanishes, $A_{\alpha\beta}=0$. This is possible because there are three degrees of freedom in both the gauge fields and the transformations.

Implementing this gauge-fixing at the level of the path integral introduces Fadeev-Popov ghosts; after all Lagrange multipliers have been integrated out, the gauge-fixed bosonic model is described by the free action:
\be\label{gfb1}
S=\frac{1}{2\pi}\int_{\Sigma} P_{\mu}\,\dbar X^{\mu}  + \sum_{a=1}^{3} b^{(a)}\,\dbar c^{(a)} \,+ S_{\mathfrak{g}} + S_{\tilde{\mathfrak{g}}}\,.
\ee
Here $c^{(1)}$ is the ghost field associated with the $P^2=0$ constraint, while $c^{(2)}$ and $c^{(3)}$ are the ghosts associated with the $X^2=0$ and $X\cdot P=0$ constraints, respectively. The $b^{(a)}$, for $a=1,2,3$ are the conjugate anti-ghosts. Note that all three ghost/anti-ghost systems are chiral (left-moving) with fermionic statistics; the ghost conformal weights are $(-1,0)$ for $c^{(1)}$, $(1,0)$ for $c^{(2)}$, and $(0,0)$ for $c^{(3)}$. The conformal weights of the anti-ghosts can be read off by requiring that $b^{(a)}\dbar c^{(a)}$ has conformal weight $(1,1)$.

Associated to this gauge fixing is a BRST operator, $Q$, made up of two contributions: an `abelian' piece given by each ghost field paired with its associated constraint, and a `non-abelian' piece composed of combinations of the ghosts and anti-ghosts which encode the structure constants of the underlying $\mathfrak{sl}(2)$ algebra. Indeed, when gauge-fixing any non-abelian constraint algebra in two dimensions, the resulting BRST charge takes the form
\be\label{naQ}
Q=\oint c^{(a)}\,T_{(a)} -\frac{1}{2} f_{abc}\,c^{(a)}\,c^{(b)}\,b^{(c)}\,,
\ee
where $\{T_{(a)}\}$ are the currents corresponding to the gauged constraints and $f_{abc}$ are their associated structure constants. In the case at hand, this leads to a BRST charge for the gauge-fixed bosonic model:
\be\label{bBRST1}
Q=\oint \frac{c^{(1)}}{2}\,P^{2} + \frac{c^{(2)}}{2}\,X^{2} + c^{(3)}\,X\cdot P - 2\left(b^{(1)}\,c^{(1)} - b^{(2)}\,c^{(2)}\right)c^{(3)} - b^{(3)}\,c^{(1)}\,c^{(2)}\,,
\ee
with all terms assumed to be normal-ordered. This operator is nilpotent, $Q^2=0$, if and only if the gauge fixing is quantum mechanically consistent (i.e., anomaly free). Obstructions to $Q^2=0$ are thus to be viewed as $\SL(2)$ anomalies associated to the algebra of constraints for the projective null cone.

The action \eqref{gfb1} has free OPEs
\be\label{OPE1}
P_{\mu}(z)\,X^{\nu}(w)\sim \frac{\delta_{\mu}^{\nu}}{z-w}\,, \qquad b^{(a)}(z)\,c^{(b)}(w)\sim \frac{\delta^{ab}}{z-w}\,,
\ee
so all anomalies can be computed exactly. Indeed, a straightforward calculation shows that
\be\label{banom1}
Q^{2} = \frac{(D-8)}{2}\left(\partial c^{(1)}\,c^{(2)}-c^{(1)}\,\partial c^{(2)} + 2 c^{(3)}\,\partial c^{(3)}\right)\,.
\ee
Remarkably, all anomalies are controlled by the dimension $D$ of the target space, and can be eliminated by fixing $D=8$. Note that the choice of $D=8$ is equivalent to fixing the `physical' dimension $d=6$, since the target lies within the projective light cone of $\C^{8}$.

At this point, the bosonic model is defined as a 2d CFT on the Riemann sphere $\Sigma\cong\CP^1$ with holomorphic central charge\footnote{Contributions to the central charge are as follows: $+2D$ from the $(P,X)$ system, $-26$ from the $(c^{(1)},b^{(1)})$ system, $-2$ from the $(c^{(2)},b^{(2)})$ system, and $-2$ from the $(c^{(3)},b^{(3)})$ system.} $\mathfrak{c}=\mathfrak{c}_{\mathfrak{g}}+\mathfrak{c}_{\tilde{\mathfrak{g}}}-14$, where $\mathfrak{c}_{\mathfrak{g}}$ is the central charge of the worldsheet current algebra $S_{\mathfrak{g}}$. However, if holomorphic (left-moving) worldsheet gravity is also gauged, then the action and BRST-charge are altered:
\be\label{gfb2}
S\rightarrow S +\frac{1}{2\pi}\int_{\Sigma} b^{(0)}\,\dbar c^{(0)}\,, \qquad Q\rightarrow Q + \oint c^{(0)}\,T + b^{(0)}\,c^{(0)}\,\partial c^{(0)}\,,
\ee
where $c^{(0)}$ is the usual fermionic reparametrization ghost of conformal weight $(-1,0)$, and $T$ is the holomorphic stress tensor
\begin{equation*}
 T=P_{\mu}\,\partial X^{\mu} - 2b^{(1)}\,\partial c^{(1)}-\partial b^{(1)}\,c^{(1)} + \partial b^{(2)}\,c^{(2)} - b^{(3)}\,\partial c^{(3)}\,.
\end{equation*}
Keeping $D=8$, there is now an additional conformal anomaly, proportional to the total central charge,\footnote{Now there is also a central charge contribution of $-26$ from the $(c^{(0)},b^{(0)})$ system.} which obstructs $Q^2=0$:
\be\label{banom2}
Q^{2}=\left(\mathfrak{c}_{\mathfrak{g}}+\mathfrak{c}_{\tilde{\mathfrak{g}}} - 40\right) \frac{c^{(0)}\,\partial^{3} c^{(0)}}{12}\,.
\ee
The conformal anomaly can thus be eliminated by choosing worldsheet current algebras with total central charge $+40$. In this case, the bosonic model can be interpreted as a (holomorphic) string theory in its own right, and (at least in principle) can be defined on a Riemann surface $\Sigma$ of \emph{any} genus. Although this places some constraints on the worldsheet current algebra, it does not fix the gauge group, since the level is still a free variable in the formula:
\begin{equation*}
 \mathfrak{c}_{\mathfrak{g}}=\frac{2 k\,\mathrm{dim}\mathfrak{g}}{c_{2}+2k}\,,
\end{equation*}
where $c_2$ is the quadratic Casimir of $\mathfrak{g}$ and $k$ is the level. For instance, supposing that $\mathfrak{g}=\tilde{\mathfrak{g}}$, the central charge can be eliminated by choosing a level $k=1$ worldsheet current algebra $\mathfrak{g}=\mathfrak{su}(21)$ or $\mathfrak{g}=\mathfrak{so}(40)$. We comment further on the issue of central charges below.

\medskip

Having discussed the gauge fixing of the bosonic model in some detail, it is easy to see that a similar story goes through for the heterotic and type II models. In the case of the heterotic model, there are two fermionic Lagrange multiplier fields $\chi^{(1)},\chi^{(2)}$ to be gauge-fixed to zero; the resulting free action on the Riemann sphere is
\be\label{gfh1}
S=\frac{1}{2\pi}\int_{\Sigma} P_{\mu}\,\dbar X^{\mu} -\frac{1}{2}\psi_{\mu}\,\dbar\psi^{\mu}  +\sum_{a=0}^{3}b^{(a)}\,\dbar c^{(a)} + \sum_{\alpha=1}^{2}\beta^{(\alpha)}\,\dbar\gamma^{(\alpha)}
\,+ S_{\mathfrak{g}}\,,
\ee
where holomorphic worldsheet gravity has also been gauge-fixed. The new bosonic ghosts $\gamma^{(1)}$ and $\gamma^{(2)}$ are associated with the constraints $\psi\cdot P=0$ and $\psi\cdot X=0$; they have conformal weights $(-\frac{1}{2},0)$ for $\gamma^{(1)}$ and $(\frac{1}{2},0)$ for $\gamma^{(2)}$. The corresponding BRST charge is given by:
\begin{multline}\label{hBRST1}
 Q=\oint c^{(0)}\,T + b^{(0)}\,c^{(0)}\,\partial c^{(0)} + \frac{c^{(1)}}{2}\,P^{2} + \frac{c^{(2)}}{2}\,X^{2} + c^{(3)}\,X\cdot P -\gamma^{(1)}\,\psi\cdot P -\gamma^{(2)}\,\psi\cdot X \\
 - 2\left(b^{(1)}\,c^{(1)} - b^{(2)}\,c^{(2)}\right)c^{(3)} - b^{(3)}\,c^{(1)}\,c^{(2)} +\gamma^{(1)}\left(c^{(2)}\,\beta^{(2)}+c^{(3)}\,\beta^{(1)}-\gamma^{(1)}\,b^{(1)}\right) \\
 -\gamma^{(2)}\left(c^{(3)}\,\beta^{(2)}+\gamma^{(2)}\,b^{(2)}+c^{(1)}\,\beta^{(1)}\right) -\gamma^{(1)}\,\gamma^{(2)}\,b^{(3)}\,.
\end{multline}
The first line can be thought of as the `abelian' piece, while the following two lines are the `non-abelian' contributions.

Using the free OPEs \eqref{OPE1} along with
\be\label{OPE2}
\beta^{(\alpha)}(z)\,\gamma^{(\beta)}(w)\sim \frac{\delta^{\alpha\beta}}{z-w}\,,
\ee
the anomalies can be computed exactly, and it follows that
\begin{multline}\label{hanom1}
 Q^{2}=(D-6)\left(\frac{\partial c^{(1)}\,c^{(2)}}{2}-\frac{c^{(1)}\,\partial c^{(2)}}{2}-\partial c^{(3)}\,c^{(3)}+\partial\gamma^{(1)}\,\gamma^{(2)}-\gamma^{(1)}\,\partial\gamma^{(2)}\right) \\
 +\frac{(D-6)}{2} \left(\partial^{2}c^{(0)}\,c^{(3)}+c^{(0)}\,\partial^{2}c^{(3)}\right)+\left(\frac{5D}{2}-46 +\mathfrak{c}_{\mathfrak{g}}\right) \frac{c^{(0)}\,\partial^{3} c^{(0)}}{12}\,.
\end{multline}
The first line contains the anomalies associated with the OSp$(2|1)$ group of gauge symmetries, while the second line contains contributions from the anomalous conformal weights of the gauge symmetries and the overall conformal anomaly. Just like the bosonic model, all anomalies related to the gauge symmetries or their conformal weights are killed by fixing the target space dimension. In the heterotic case, the target dimension must be $D=6$, or equivalently, the `physical' dimension is fixed to $d=4$. The remaining conformal anomaly can then be eliminated if desired by fixing $\mathfrak{c}_{\mathfrak{g}}=31$. For example, level one worldsheet current algebras for $\mathfrak{g}=\mathfrak{su}(32)$ and $\mathfrak{g}=\mathfrak{so}(62)$ satisfy this criterion. 

In the case of the type II model, all Lagrange multiplier fields can also be gauge-fixed to zero using the freedoms \eqref{IItrans1}--\eqref{IItrans3}. Once more, the result is a free action on $\Sigma$:
\begin{multline}\label{gfII1}
 S= \frac{1}{2\pi}\int_{\Sigma} P_{\mu}\,\dbar X^{\mu} -\frac{1}{2}\psi_{\mu}\,\dbar\psi^{\mu} -\frac{1}{2}\tpsi_{\mu}\,\dbar\tpsi^{\mu} +\sum_{a=0}^{3}b^{(0)}\,\dbar c^{(0)} \\
 + \sum_{\alpha=1}^{2}\left(\beta^{(\alpha)}\,\dbar\gamma^{(\alpha)}+\tilde{\beta}^{(\alpha)}\,\dbar\tilde{\gamma}^{(\alpha)}\right) + \lambda\,\dbar\eta\,. 
\end{multline}
The bosonic ghosts $\tilde{\gamma}^{(1)}$, $\tilde{\gamma}^{(2)}$ have the same conformal weights as their un-tilded cousins and are associated with the constraints $\tpsi\cdot P=0$ and $\tpsi\cdot X=0$, respectively. There is also a new fermionic ghost $\eta$, of conformal weight $(0,0)$, associated with the nilpotent bosonic constraint $\psi\cdot\tpsi=0$. To arrive at this free action, four bosonic and four fermionic Lagrange multipliers have been gauge-fixed to zero, so the resulting BRST charge has a rather lengthy expression:
\begin{multline}\label{IIBRST1}
 Q=\oint c^{(0)}\,T + b^{(0)}\,c^{(0)}\,\partial c^{(0)} + \frac{c^{(1)}}{2}\,P^{2} + \frac{c^{(2)}}{2}\,X^{2} + c^{(3)}\,X\cdot P -\gamma^{(1)}\,\psi\cdot P -\gamma^{(2)}\,\psi\cdot X \\
 -\tilde{\gamma}^{(1)}\,\tpsi\cdot P - \tilde{\gamma}^{(2)}\,\tpsi\cdot X  -\eta\,\psi\cdot\tpsi - 2\left(b^{(1)}\,c^{(1)} - b^{(2)}\,c^{(2)}\right)c^{(3)} - b^{(3)}\,c^{(1)}\,c^{(2)} \\
 +\gamma^{(1)}\left(c^{(2)}\,\beta^{(2)}+c^{(3)}\,\beta^{(1)}-\gamma^{(1)}\,b^{(1)}\right) -\gamma^{(2)}\left(c^{(3)}\,\beta^{(2)}+\gamma^{(2)}\,b^{(2)}+c^{(1)}\,\beta^{(1)}\right) -\gamma^{(1)}\,\gamma^{(2)}\,b^{(3)} \\
 +\tilde{\gamma}^{(1)}\left(c^{(2)}\,\tilde{\beta}^{(2)}+c^{(3)}\,\tilde{\beta}^{(1)}-\tilde{\gamma}^{(1)}\,b^{(1)}\right) -\tilde{\gamma}^{(2)}\left(c^{(3)}\,\tilde{\beta}^{(2)}+\tilde{\gamma}^{(2)}\,b^{(2)}+c^{(1)}\,\tilde{\beta}^{(1)}\right) -\tilde{\gamma}^{(1)}\,\tilde{\gamma}^{(2)}\,b^{(3)} \\
 +\left(\gamma^{(1)}\,\tilde{\gamma}^{(2)}-\gamma^{(2)}\,\tilde{\gamma}^{(1)}\right)\lambda +\eta\left(\gamma^{(1)}\,\tilde{\beta}^{(1)}+\gamma^{(2)}\,\tilde{\beta}^{(2)} -\tilde{\gamma}^{(1)}\,\beta^{(1)}-\tilde{\gamma}^{(2)}\,\beta^{(2)}\right)\,,
\end{multline}
where the final line arises from the `non-abelian' interplay between the fermionic currents (tilded and un-tilded) and the $\psi\cdot\tpsi$ current.

Despite the somewhat opaque form of \eqref{IIBRST1}, $Q^2$ can still be computed explicitly using the free OPEs of \eqref{gfII1} leading to:
\begin{multline}\label{IIanom1}
 Q^2 = (D-4)\left(\frac{\partial c^{(1)}\,c^{(2)}}{2}-\frac{c^{(1)}\,\partial c^{(2)}}{2}-\partial c^{(3)}\,c^{(3)}+\partial\gamma^{(1)}\,\gamma^{(2)}-\gamma^{(1)}\,\partial\gamma^{(2)} +\partial\tilde{\gamma}^{(1)}\,\tilde{\gamma}^{(2)}\right.\\ 
 -\tilde{\gamma}^{(1)}\,\partial\tilde{\gamma}^{(2)}-\partial\eta\,\eta\bigg) +\frac{(D-4)}{2} \left(\partial^{2}c^{(0)}\,c^{(3)}+c^{(0)}\,\partial^{2}c^{(3)}\right) +\left(3D-34\right) \frac{c^{(0)}\,\partial^{3} c^{(0)}}{12}\,.
\end{multline}
Remarkably, in the type II model as in the bosonic and heterotic models, the $\SL(2|1)$ gauge anomalies and the mixed gauge/conformal anomaly are all controlled by the target dimension. These anomalies are eliminated for the type II model by fixing $D=4$, or the `physical' dimension $d=2$. With this choice the conformal anomaly, or holomorphic central charge, of the model becomes $\mathfrak{c}=-22$. So long as $\Sigma\cong\CP^1$ this is not a dangerous anomaly.

\medskip

So to reiterate the salient points: each of the three models (bosonic, heterotic, type II) can be gauge-fixed so that the non-dynamical gauge field vanishes. In each case, the gauge anomalies (i.e., those associated with the symmetries generated by the constraints) are completely controlled by the target space dimension. For the bosonic model, the gauge anomalies are killed for $D=8$ or $d=6$; for the heterotic model it is $D=6$ or $d=4$; and for the type II model it is $D=4$ or $d=2$. Since the array of ghost fields appearing in these models is rather extensive, we list all the ghost fields and their conformal weights for the type II model here for future reference; quantum numbers of the anti-ghost fields are easily deduced from these. 
\begin{center}
 \begin{tabular}{||c c c c||} 
 \hline
 Ghost & Associated Constraint & Statistics & Conformal Weight \\ [0.5ex] 
 \hline\hline
 $c^{(0)}$ & $T$ & Fermionic & $(-1,0)$ \\ 
 \hline
 $c^{(1)}$ & $P^2$ & Fermionic & $(-1,0)$ \\
 \hline
 $c^{(2)}$ & $X^2$ & Fermionic & $(1,0)$ \\
 \hline
 $c^{(3)}$ & $X\cdot P$ & Fermionic & $(0,0)$ \\
 \hline
 $\gamma^{(1)},\;\tilde{\gamma}^{(1)}$ & $\psi\cdot P,\; \tpsi\cdot P$ & Bosonic & $(-\frac{1}{2},0)$ \\
 \hline
 $\gamma^{(2)},\;\tilde{\gamma}^{(2)}$ & $\psi\cdot X,\; \tpsi\cdot X$ & Bosonic & $(\frac{1}{2},0)$ \\
 \hline
 $\eta$ & $\psi\cdot\tpsi$ & Fermionic & $(0,0)$ \\ [1ex]
 \hline
\end{tabular}
\end{center}


\subsection{Vertex operators}

The specific target space dimensions selected to eliminate the gauge anomalies of the models certainly suggest a physical space-time interpretation. For instance, the bosonic model has a target space which is a natural arena for describing $d=6$ conformal field theories. Likewise, the target spaces of the heterotic and type II models are the natural arenas for $d=4$ and $d=2$ conformal field theories, respectively. To establish the extent to which these critical dimensions (and their relationship to classical space-time conformal invariance) are meaningful, it makes sense to investigate the spectra of the three theories. In each case, we have $Q^2=0$ (at least, up to a conformal anomaly which is irrelevant at genus zero), so the BRST cohomology is well-defined and vertex operators are composite operators $V$ which are BRST closed: $QV=0$.

Generally speaking, an un-integrated vertex operator in a 2d CFT should have positive ghost number and bosonic statistics. We also restrict our attention to operators with vanishing conformal weight on $\Sigma$. In the case of the bosonic model, one particular ansatz which satisfies these properties is
\be\label{VOa1}
V=c^{(0)}\,c^{(1)}\,j^{\sa}\,\tilde{j}^{\sta}\,f(X)\,,
\ee
where $j^{\sa}$ is the conformal weight $(1,0)$ current associated with the worldsheet current algebra action $S_{\mathfrak{g}}$, taking values in the adjoint of $\mathfrak{g}$: $\sa=1,\ldots,\mathrm{dim}\,\mathfrak{g}$. The current $\tilde{j}^{\sta}$ is the one associated with the second worldsheet current algebra, $S_{\mathfrak{\tilde{g}}}$. The OPEs of these currents obey
\be\label{CA}
j^{\sa}(z)\,j^{\mathsf{b}}(w)\sim \frac{k\,\delta^{\sa \mathsf{b}}}{(z-w)^2}+\im\frac{f^{\sa \mathsf{b}\mathsf{c}}\,j^{\mathsf{c}}(w)}{z-w}\,,
\ee
where $k$ is the level of the worldsheet current algebra, and $\delta^{\sa \mathsf{b}}$, $f^{\sa \mathsf{b}\mathsf{c}}$ are the Killing form and structure constants of $\mathfrak{g}$, respectively.

The ansatz \eqref{VOa1} has vanishing conformal weight (i.e., is a scalar operator on $\Sigma$) for any choice of function $f(X)$, so the BRST-closure condition should place some restrictions on this otherwise freely specified function. Indeed, it is easy to see that
\be\label{bVO1}
QV=j^{\sa}\,\tilde{j}^{\sta}\left[\frac{c^{(0)}c^{(1)}\partial c^{(1)}}{2}\frac{\partial^{2} f}{\partial X^{\mu}\partial X_{\mu}} +c^{(0)}c^{(1)}c^{(3)}\left(X\cdot\frac{\partial f}{\partial X}+2\,f\right)\right]\,.
\ee
That is, $V$ lies in the BRST cohomology if $f$ is a solution to the wave equation in the $D=8$ dimensional embedding space \emph{and} has homogeneity $-2$ in $X$. These conditions are precisely those which are required for $f(X)$ to represent a conformal weight $\Delta=2$ scalar field in $d=6$ (e.g., \cite{Dirac:1936fq,Penedones:2010ue}).

Specific solutions to the $QV=0$ conditions can be provided by taking
\be\label{bf1}
f^{[p]}(X)=\frac{\cP_{\mu_{1}\cdots\mu_{p}} X^{\mu_1}\cdots X^{\mu_p}}{(k\cdot X)^{2+p}}\,,
\ee
with $k^{2}=0$ and $\cP_{\mu_1 \cdots\mu_p}=\cP_{(\mu_1 \cdots\mu_p )}$ obeying $k^{\mu_1}\cP_{\mu_1 \cdots\mu_p}=0$. Here the `momentum' $k_{\mu}$ should actually be viewed as a point in $d=6$ Minkowski space (since $k^2=0$ indicates that it lies on the $D=8$ null cone); and \eqref{bf1} plays a role analogous to the plane wave factor $e^{\im \kappa\cdot x}$ of ambitwistor string vertex operators. Indeed, $f^{[0]}$ is simply the Green's function for a $\Delta=2$ scalar field; $\{f^{[p]}\}$ for all $p>0$ form the tower of conformal descendants of this basic scalar field.

So the family of vertex operators
\be\label{bVO2}
V^{[p]}=c^{(0)}\,c^{(1)}\,j^{\sa}\,\tilde{j}^{\sta}\,f^{[p]}(X)\,,
\ee
represent single field insertions of a dimension $\Delta=2$ scalar field ($p=0$) and all of its conformal descendants in $d=6$ space-time. The presence of the two gauge currents means that this family of operators is valued in the bi-adjoint of the gauge algebras $\mathfrak{g}$, $\tilde{\mathfrak{g}}$. In other words, the BRST cohomology of the bosonic model contains vertex operators representing single field insertions of a bi-adjoint scalar field in $d=6$, along with all of this bi-adjoint scalar's conformal descendants. This is no coincidence: $d=6$ is precisely the dimension for which the cubic bi-adjoint scalar theory is (classically) conformally invariant. This is a theory of scalars $\phi^{\sa\sta}$ with a cubic interaction term $f^{\sa \mathsf{b} \mathsf{c}}\tilde{f}^{\sta \tilde{\mathsf{b}} \tilde{\mathsf{c}}}\phi^{\sa\sta}\phi^{\mathsf{b}\tilde{\mathsf{b}}}\phi^{\mathsf{c}\tilde{\mathsf{c}}}$.

\medskip

Analogous families of vertex operators lie within the BRST cohomology of the heterotic and type II models. In the heterotic model, consider un-integrated vertex operators of the form:
\be\label{hVO1}
V=c^{(0)}\,c^{(1)}\,\delta(\gamma^{(1)})\,j^{\sa}\,\psi^{\mu}\,A_{\mu}(X)\,,
\ee
for some vector $A_\mu$ which is a function only of $X$. Here $\delta(\gamma^{(1)})$ fixes the value of the bosonic ghost $\gamma^{(1)}$ to vanish at the vertex operator insertion; since $\gamma^{(1)}$ carries conformal weight $(-\frac{1}{2},0)$, the delta function has conformal weight $(\frac{1}{2},0)$. This ansatz has vanishing conformal weight, and the condition for $Q$-closure is
\begin{multline}\label{hVO2}
 QV=\delta(\gamma^{(1)})\,j^{\sa}\left[\frac{c^{(0)}c^{(1)}\partial c^{(1)}}{2}\psi^{\mu} \frac{\partial^{2} A_{\mu}}{\partial X^{\nu}\partial X_{\nu}}-c^{(0)}c^{(1)}\gamma^{(2)} X\cdot A \right. \\
 \left.+c^{(0)}c^{(1)}c^{(3)}\psi^{\mu}\left(X\cdot\frac{\partial A_{\mu}}{\partial X}+A_{\mu}\right) -c^{(0)}c^{(1)}\partial\gamma^{(1)} \frac{\partial A_{\mu}}{\partial X^{\mu}}\right]=0\,.
\end{multline}
This places four conditions on the vector $A_{\mu}$; as we show in appendix \ref{appendix}, these precisely imply that $A_{\mu}$ descends to a $d=4$ gauge field. Thus, vertex operators of the form \eqref{hVO1} describe single field insertions of adjoint-valued gauge fields in $d=4$ along with all conformal descendants. Of course, this is consistent with $d=4$, where Yang-Mills theory is a (classical) conformal field theory.

Finally, the same basic ansatz can be extended to the type II model, where it becomes
\be\label{2VO1}
V=c^{(0)}\,c^{(1)}\,\delta(\gamma^{(1)})\,\delta(\tilde{\gamma}^{(1)})\,\psi^{\mu}\,\tilde{\psi}^{\nu}\,h_{\mu\nu}(X)\,,
\ee
where the free data is now a rank-two tensor $h_{\mu\nu}$ depending only on $X$. The action of $Q$ on this ansatz can again be calculated explicitly, leading to the conditions
\begin{multline}\label{2VO2}
 QV=c^{(0)}\,c^{(1)}\,\delta(\gamma^{(1)})\,\delta(\tilde{\gamma}^{(1)})\left[\frac{\partial c^{(1)}}{2}\psi^{\mu}\tpsi^{\nu}\frac{\partial^{2} h_{\mu\nu}}{\partial X^{\sigma}\partial X_{\sigma}}-\gamma^{(2)}\tpsi^{\nu} X^{\mu}h_{\mu\nu}+\tilde{\gamma}^{(2)}\psi^{\mu} X^{\nu}h_{\mu\nu} \right. \\
  -\left(\partial\gamma^{(1)}\tpsi^{\nu}\frac{\partial}{\partial X_{\mu}}-\partial\tilde{\gamma}^{(1)}\psi^{\mu}\frac{\partial}{\partial X_{\nu}}\right) h_{\mu\nu} + \eta \left(\psi^{\nu}\psi^{\mu}+\tpsi^{\mu}\tpsi^{\nu}\right)h_{\mu\nu} \\
 \left.+\partial\eta\,h^{\mu}_{\mu}+ c^{(3)}\,X\cdot\frac{\partial h_{\mu\nu}}{\partial X}\right]=0\,.
\end{multline}
As shown in appendix \ref{appendix}, these condition imply that $h_{\mu\nu}$ descends to a $d=2$ ``graviton'', i.e. a symmetric traceless tensor satisfying the Einstein equation. In other words, vertex operators of the form \eqref{2VO1} and subject to the conditions \eqref{2VO2} describe single field insertions of metric perturbations in $d=2$, along with all of their conformal descendants. As in the bosonic and heterotic models, this is consistent with the fact that gravity is a classical conformal field theory in two dimensions. In fact, it is actually a topological theory, and hence all correlators of these vertex operators turn out to be zero, as we will see in section~\ref{3pt}.

In the heterotic and type II models there are also `Ramond-sector' versions of \eqref{hVO1} and \eqref{2VO1} which correspond to single field insertions of gaugino and gravitino fields (and their conformal descendants) respectively. This is a strong indication that we are actually looking at \emph{super} conformal field theories in $d$ dimensions, which is to be expected from the structure of the phase space itself. For the remainder of this paper, we restrict our attention to the `NS-sector' operators, leaving the question of space-time supersymmetry to future investigations.

\medskip

While it is encouraging to see that single field insertions for these three well-known CFTs ($d=6$ biadjoint cubic scalars, $d=4$ gauge theory and $d=2$ gravity) lie within the BRST cohomology of our models, it is a far cry from proving that the operator spectra of the models coincide with the space-time CFT spectra of local operators. Indeed, the space-time CFTs have a wide array of composite operators with a variety of conformal dimensions, and such operators are not captured by the ans\"atze deployed above. This is because any vertex operator ansatz which is not proportional to the $c^{(3)}$ ghost will result in some fixed homogeneity constraint, which corresponds to fixing the space-time conformal dimension.

Conversely, it is equally clear that the ans\"atze do not account for all of the vertex operators in the BRST cohomology of the models. The bosonic and heterotic models have additional un-wanted vertex operators of the sort:
\be\label{diffanom}
 c^{(0)}\,c^{(1)}\,P_{\mu}\,P_{\nu}\, a^{\mu\nu}\,, \qquad c^{(0)}\,c^{(1)}\,\delta(\gamma^{(1)})\,P_{\mu}\,\psi^{\nu} \,a_{\nu}^{\mu}(X)\,,
\ee
which are $Q$-closed under some basic assumptions on the tensor $a_{\mu\nu}$. These vertex operators describe gravitational modes on the $d$-dimensional space-time. They are, however, un-physical because space-time diffeomorphism invariance is not built into the underlying bosonic or heterotic models on $\Sigma$. Operators of this type do not appear in the type II model, but in the bosonic and heterotic models they must be thrown away by hand. As long as $\Sigma\cong\CP^1$, such a truncation can be shown to be consistent.

However, in all three models there are many other vertex operators in the BRST cohomology which differ substantially from the subsector corresponding to our ans\"atze. For example, in the bosonic model, it is easy to see that a vertex operator of the form
\be\label{altVO}
V=c^{(0)}\,c^{(1)}\,c^{(2)}\,c^{(3)}\,\frac{j^{\sa}\,\tilde{j}^{\sta}}{\ell\cdot P}\,\e^{i k\cdot X}\,,
\ee
obeys $QV=0$ so long as $\ell^2=0=k^2$ and $k\cdot\ell=0$. The space-time interpretation of such an operator is not clear: the free data ($\ell^{\mu}$ and $k^{\mu}$) are two null separated points in $d=6$ Minkowski space, and \eqref{altVO} has no definite homogeneity in $X$ -- and hence no well-defined conformal dimension. Similar operators also appear in the BRST cohomology of the heterotic and type II models.

Optimistically, one might hope that these two issues could resolve one another. That is, the missing composite operators of the space-time CFT spectra could be encoded by other vertex operators on $\Sigma$ such as \eqref{altVO}, superpositions thereof, or something else entirely. In the absence of a concrete argument, we simply state: in each of the three models, the BRST cohomology includes all single field insertions (with conformal descendants) for a classical space-time CFT in the appropriate dimension. This correspondence is summarized in the table:
\begin{center}
 \begin{tabular}{||c c c c||} 
 \hline
 Model & Critical dimension & Space-time CFT & Single field \\ [0.5ex] 
 \hline\hline
 Bosonic & $d=6$ & Biadjoint $\phi^3$ & Scalars \\ 
 \hline
 Heterotic & $d=4$ & Yang-Mills & Gauge fields \\
 \hline
 Type II & $d=2$ & Gravity & Gravitons \\ [1ex]
 \hline
\end{tabular}
\end{center}
We leave the questions regarding the remainder of the spectra (on both $\Sigma$ and space-time) to future work.


\subsection{Anomalies and higher genus}

At this point it seems appropriate to comment on the issue of central charges, which we have alluded to above. In the bosonic and heterotic models, the conformal anomaly can be eliminated simultaneously with the gauge anomalies by a judicious choice of the central charges of the current algebras. There is no such freedom in the type II model. On the other hand, the bosonic and heterotic models possess a diffeomorphism anomaly, which manifests itself in the un-wanted gravitational vertex operators \eqref{diffanom}. 

This can be seen by analogy with the ambitwistor string counterparts of these models~\cite{Mason:2013sva,Adamo:2014wea}. In fact, after gauge-fixing the non-dynamical gauge field to zero, our models have the same free $(P,X)$ action as their ambitwistor string counterparts, and differ only in their ghost systems (and BRST charge). To `fix' the bosonic or heterotic models, the unwanted vertex operators should be discarded by hand. But this prescription is only consistent at genus zero, since the bad states will spoil the models at higher genus.

Although the type II model is also restricted to genus zero, this is only on account of its non-vanishing central charge. The diffeomorphism anomaly that plagues the bosonic and heterotic models is absent in the type II ambitwistor string \cite{Adamo:2014wea}, as the anomalies from the $PX$ system and from the fermionic systems $\psi$ and $\tilde\psi$ cancel each other. This is a delicate cancellation (related to general properties of curved $\beta\gamma$-systems~\cite{Nekrasov:2005wg}) and seems fairly robust and unique: the type II ambitwistor string is the only known ambitwistor model which avoids the diffeomorphism anomaly.

For these reasons, none of the three models that we presented can be considered a fully fledged \emph{string} theory. There is, however, an alternative path. Recent work \cite{Geyer:2015bja,Geyer:2015jch,Geyer:2016wjx} on ambitwistor strings at loop level suggests that there is a new breed of worldsheet models whose loop expansion is not a genus expansion, as in a proper string theory, but rather an expansion in nodes (pairs of identified points) of the Riemann sphere -- the loop momenta run through these nodes. If such models can be properly defined at any loop order -- which is not established -- then the consistency conditions will likely be weaker than those of a string theory. After all, we are only meant to be describing theories of particles, albeit in a worldsheet formulation.


\subsection{Gauge-fixing and the ambitwistor string}

Thus far, we have only considered the particular gauge fixing that sets all Lagrange multipliers in the initial action to zero. For the bosonic model, this corresponds to setting the $\SL(2)$ gauge field to zero. However, we are of course free to pick other gauges. One particularly interesting choice allows us to recover the ambitwistor string from our model. Instead of setting $e^{(2)}$ and $e^{(3)}$ to zero, we choose $X^-$ to solve $X^2=0$, and choose $P^-$ to solve $X\cdot P=0$, obtaining
\be
X^\mu=X^+(1,\,x^2,\,x^a)\,, \qquad P^\mu=(P^+,\,2 p'\cdot x-P^+x^2,\,p'^a)\,.
\ee
Substituting into the action and performing field redefinitions, we get
\be
S=\int_\Sigma \left(p_a\, \dbar x^a-\frac{e}{2}\, p^2\right)\,, \qquad p^a=X^+(p'^a-P^+x^a)\,, \quad
e=\frac{e^{(1)}}{(X^+)^2}\,,
\ee
which we recognize as the ambitwistor string action for a $d$-dimensional target space (i.e., two dimensions less than what we started with). It is important to emphasize that the absence of anomalies for the $\SL(2)$ gauge field is a gauge invariant statement, which implies that we can only recover the ambitwistor string in the correct space-time dimension: $d=D-2=6$ for the biadjoint scalar.

For the heterotic model, the story follows similarly, with the additional choice of $\psi^-$ to solve $X\cdot\psi=0$,
\be
\psi^\mu=(\psi^+,\,2 \varphi'\cdot x-\psi^+x^2,\,\varphi'^a)\,,
\ee
leading to the action
\be
S=\int_\Sigma \left(p\cdot \dbar x-\frac{1}{2}\,\varphi\cdot\dbar\varphi-\frac{e}{2}\, p^2+\chi\, p\cdot\varphi\right)+S_{\mathfrak{g}}\,, \quad \varphi^a=\varphi'^a-\psi^+x^a\,, \quad
\chi=\frac{\chi^{(1)}}{X^+}\,,
\ee
which is the heterotic ambitwistor string action, now derived for $d=4$ dimensions.

For the type II model, we use the same choices to solve $X\cdot\psi=0$ and $X\cdot\tilde\psi=0$, but the action obtained differs from the type II ambitwistor action in $d=2$ dimensions. The difference is that the constraint term \,$\int_\Sigma\; \xi \, \varphi\cdot\tilde{\varphi}$\, remains. This constraint changes the BRST charge and restricts the vertex operator \eqref{2VO1} to a traceless symmetric $h_{\mu\nu}$, representing a spin-2 field. By contrast, the analogous plane-wave vertex operator of the ambitwistor string describes a general rank-2 tensor, with graviton, dilaton and B-field components.

In the bosonic and heterotic cases, we saw that our original actions are actually completely equivalent to the ambitwistor strings in a specific target space dimension. Na\"ively this may seem surprising, since the latter describe on-shell scattering amplitudes while the former are supposed to compute correlation functions. There is of course no fundamental incompatibility, since these actions only describe weakly coupled theories. For these the LSZ prescription works, so it is clear that one may move from correlators to scattering amplitudes via the Fourier transform. 

At the level of vertex operators and the physical state space of the theories, there must be an equivalence. For instance, in the case of the biadjoint cubic scalar theory, the cohomology contains operators of the form
\be
V\sim \frac{\delta(X^2)}{(k\cdot X)^2}, \qquad k^2=0\,,
\ee
with the delta function provided by the path integral measure. In the `ambitwistor gauge fixing', this vertex operator looks like
\be
\tilde V\sim\frac{1}{(y-x)^4}\,,
\ee
where we have set $k=(1,y^2,y^a)$. In this expression $y^a$ are fixed parameters whereas $x^a$ are fields on $\Sigma$. Since this takes the form of a free propagator, we can trivially write it as an explicit superposition of plane-wave vertex operators of the ambitwistor string, namely:
\be
\tilde V\sim\int d^6 q\, \delta(q^2) \left( \theta(q^0)\,\e^{\im q\cdot(x-y)} + \theta(-q^0) \e^{\im q \cdot (y-x)}\right)\,.
\ee

Although the two theories are in this sense equivalent, this does not mean that they are equally well-suited for computing correlators and scattering amplitudes. The whole point is that we {\em do not} want to compute Fourier transforms to obtain correlators, much as we do not want to compute off-shell correlators and LSZ reduce them to obtain scattering amplitudes. In particular, the similarity of our theories with an ambitwistor string in $D$ dimensions (i.e., with extra constraints) suggests there should be a simple prescription for computing correlators directly. As a first step in this direction, we now turn to the computation of three point functions.


\section{A Prescription for Three-point Functions}
\label{3pt}

The models discussed above encode classical conformal invariance on space-time for three well-known field theories through their critical target dimensions and their vertex operator spectra. That this classical property of the space-time theories emerges from \emph{quantum} properties of the 2d models is particularly remarkable. A natural question is then: what do the 2d correlation functions of vertex operators in these models correspond to? An optimistic conjecture is that correlation functions on $\Sigma$ are related to tree-level contributions to correlation functions in the space-time CFTs.

Providing an affirmative answer to this conjecture is beyond the scope of this paper, but we present some promising evidence in this direction by considering 3-point correlators of those vertex operators which represent single field insertions. In the case of the bosonic model, the simplest example of such a correlation function is given by:
\be\label{bcor1}
\left\la \prod_{i=1}^{3} V^{[0]}(z_i)\right\ra\,,
\ee
where $V^{[0]}$ are the vertex operators \eqref{bVO2} inserted at positions $\{z_{1},z_{2},z_{3}\}$ on the Riemann sphere $\Sigma$, and the brackets $\la \cdots \ra$ stand for the correlator evaluated in the path integral of the gauge-fixed action \eqref{gfb2}. 

There are various ghost zero modes which must be saturated in this correlation function in order for the path integral to be non-zero. Since all the fields are chiral, first-order systems on $\Sigma$, their zero modes can be counted in the usual way using the Riemann-Roch theorem. In particular, a field of conformal weight $(w,0)$ is a section of the $w^{\mathrm{th}}$-power of the holomorphic cotangent bundle (or canonical bundle), $K_{\Sigma}$; its zero modes are the global holomorphic sections of this bundle: $H^{0}(\Sigma, K^{w}_{\Sigma})$.\footnote{Recall that a worldsheet field of conformal weight $(w,0)$ can be viewed as an element of $\Omega^{0}(\Sigma, K_{\Sigma}^{w})$, the space of sections of $K^{w}_{\Sigma}$. Locally, such sections have the form $f=f(z,\bar{z})\,(\d z)^{w}$. $H^{0}(\Sigma, K^{w}_{\Sigma})$ denotes the space of all such sections which are globally holomorphic, $\dbar f=0$.} On the Riemann sphere, $\Sigma\cong\CP^1$, there are no global holomorphic sections if $w>0$ since $K_{\Sigma}\cong\cO(-2)$ has negative degree. If $w\leq 0$, then the Riemann-Roch formula gives the number of zero modes:
\begin{equation*}
 h^{0}(\Sigma, K^{w}_{\Sigma})= 1-2w\,.
\end{equation*}
For example, the ghosts $c^{(0)}$ and $c^{(1)}$ both have conformal weight $(-1,0)$, or equivalently, are sections of the holomorphic tangent bundle, $T_{\Sigma}\cong K^{-1}_{\Sigma}$. The Riemann-Roch formula immediately tells us that they have three zero modes each.

Similarly, it follows that $c^{(3)}$ has a single zero mode, and there are no $c^{(2)}$ zero modes. In addition, there is a single anti-ghost zero mode for $b^{(2)}$, which is associated to the K\"ahler form on $\Sigma$. In other words, the zero mode of $b^{(2)}$ is naturally paired with  
\be\label{kahler}
\omega_{0}= \frac{\d z\wedge\d\bar{z}}{(1+|z|^{2})^2}\,,
\ee
the canonical section of $H^{1}(\Sigma, K_{\Sigma})$.

In the presence of vertex operators, the gauge-fixing of section \ref{QMod} becomes non-trivial. In the ambitwistor string, where the non-dynamical gauge field is abelian, this non-triviality took the form of a cohomological obstruction to achieving the $e^{(1)}=0$ gauge -- it is precisely this obstruction which leads to the scattering equations~\cite{Adamo:2013tsa,Ohmori:2015sha}. For the $\SL(2)$ non-dynamical gauge field of the bosonic model, the obstruction is again cohomological because the symmetries of the action are nilpotent: the holomorphic $\SL(2)$ connection is flat. With vertex operators inserted at marked points on $\Sigma$, we expect the appearance of certain delta functions in the path integral which fix the constraints dual to the gauge field at each marked point. 

The role of these delta functions is to enforce the $\SL(2)$ triplet of constraints $P^{2}=X^{2}=X\cdot P=0$ within the gauge-fixed path integral. The gauge-fixing itself is not sufficient to fix all of the gauge-field moduli because of the cohomological obstructions; the delta functions fix the remaining degrees of freedom. This is similar to what occurs in the study of dynamical gauge fields on the marked Riemann sphere, where the gauge-fixed path integral contains insertions of delta functions fixing the holonomy of the gauge connection to lie within a prescribed conjugacy class at each marked point~\cite{Witten:1991we}.

Consider the component of the $\SL(2)$ gauge field $e^{(a)}$ for $a=1,2,3$. The obstruction for this portion of the gauge field is the part of $e^{(a)}$ which cannot be set to zero by a $\SL(2)$ gauge transformation which vanishes at the vertex operator insertion points. If $\bar{D}=\dbar+A$ is the covariant derivative with respect to the $\SL(2)$ gauge field $A$, then under a gauge transformation $M$, $\delta e^{(a)}=\bar{D}M^{(a)}$ by \eqref{nab1}. So the obstruction is the part of $e^{(a)}$ which is not $\bar{D}$-exact. Since $\bar{D}^2=0$, this is precisely the cohomology group $H^{1}(\Sigma, K_{\Sigma}^{w_a}(-z_i))$, where $w_a$ is the holomorphic conformal weight of $e^{(a)}$ and $K_{\Sigma}(-z_i)$ is the bundle of holomorphic 1-forms on $\Sigma$ with zeros at the vertex operator insertion points $\{z_i\}$.

The Riemann-Roch theorem can be used to compute the number of obstructions for each gauge field -- and hence the number of additional delta function insertions in the path integral. For $n$ vertex operator insertions and $\Sigma\cong\CP^1$, there are $n-3$ obstructions to setting $e^{(1)}$ to zero, $n+1$ obstructions to setting $e^{(2)}$ to zero, and $n-1$ obstructions to setting $e^{(3)}$ to zero. This means that the path integral must contain $n-3$ delta functions fixing $P^2$, $n+1$ delta function fixing $X^2$ and $n-1$ delta functions fixing $X\cdot P$, or else the $\SL(2)$ triplet of constraints will not be satisfied.

\medskip

The synthesis of all these observations is the following prescription for the 3-point correlator:
\begin{multline}\label{bcor2}
 \left\la \prod_{i=1}^{3} V^{[0]}(z_i)\right\ra:=\int[\D\cF]\,c^{(3)}(z_{0})\,\left(b^{(2)}|\omega_{0}\right)\,\delta(X^{2}(z_0))\,\prod_{j=1}^{3}\delta\!\left(\log\mathrm{Hol}_{z_j} X^{2}\right) \\
 \prod_{k=1}^{2}\delta\!\left(\mathrm{Res}_{z=z_k} X\cdot P\right)\,\prod_{i=1}^{3} V^{[0]}(z_i)\;\e^{-S}\,,
\end{multline}
where $[\D\cF]$ denotes the path integral measure over the various fields. The ingredients are as follows: a single insertion of $c^{(3)}$ at an arbitrary point $z_{0}\in\Sigma$ saturates the zero mode integral over the fermionic ghosts; the pairing
\begin{equation*}
 \left(b^{(2)}|\omega_0\right):=\int_{\Sigma} b^{(2)}\,\omega_0\,,
\end{equation*}
saturates the path integral in the fermionic anti-ghosts and also indicates that one of the delta function constraints on $X^2$ is redundant. The first delta function sets to zero the function $X^2$ at an arbitrary point on $\Sigma$, chosen to coincide with $z_0$ without loss of generality. For any function $f$ on $\Sigma$, $\log\mathrm{Hol}_{z_j} f(z)$ stands for the phase of the holonomy of that function around the marked point $z_j$; \eqref{bcor2} contains three delta functions which restrict this phase to be zero. Likewise, $\mathrm{Res}_{z=z_{k}}$ stands for the residue map at $z_k$; if $F(z)$ is a $(1,0)$-form on $\Sigma$ which looks locally like $F(z)=f(z_k) (z-z_k)^{-1}$ then the residue map is simply
\begin{equation*}
 \mathrm{Res}_{z=z_k} F(z):= \frac{1}{2\pi \im}\int_{\Sigma} F(z)\,(z-z_k)\,\dbar\left(\frac{1}{z-z_k}\right)=f(z_k)\,.
\end{equation*}
The prescription contains two delta functions which constraint the residue of $X\cdot P$ to vanish at two of the marked points. Finally, \eqref{bcor2} contains the three vertex operator insertions and is weighted by the exponential of the action.

To evaluate the correlator in this prescription, we must perform the path integral over the various 2d CFT systems in \eqref{bcor2}. The only contribution to the path integral over fermionic ghosts comes from zero modes; these are well-known:
\begin{equation*}
 \left\la c^{(0)}(z_1)\,c^{(0)}(z_2)\,c^{(0)}(z_3)\right\ra_{b^{(0)}\dbar c^{(0)}} = \frac{(z_1-z_2) (z_2-z_3) (z_3-z_1)}{\d z_1\,\d z_2\, \d z_3}=\frac{1}{\mathrm{vol}\,\SL(2,\C)}\,,
\end{equation*}
\begin{equation*}
 \left\la c^{(1)}(z_1)\,c^{(1)}(z_2)\,c^{(1)}(z_3)\right\ra_{b^{(1)}\dbar c^{(1)}} = \frac{(z_1-z_2) (z_2-z_3) (z_3-z_1)}{\d z_1\,\d z_2\, \d z_3}=\frac{1}{\mathrm{vol}\,\SL(2,\C)}\,,
\end{equation*}
\begin{equation*}
 \left\la c^{(3)}(z_0)\right\ra_{b^{(3)}\dbar c^{(3)}} =\frac{1}{\mathrm{vol}\,\C^{*}}\,.
\end{equation*}
It is also easy to evaluate the portion of the correlator sourced by the worldsheet current algebras using \eqref{CA}:
\begin{equation*}
 \left\la j^{\sa_1}(z_1)\,j^{\sa_2}(z_2)\,j^{\sa_3}(z_3)\right\ra_{S_\mathfrak{g}}=f^{\sa_1 \sa_2 \sa_3}\frac{\d z_1\, \d z_2\, \d z_3}{(z_1-z_2) (z_2-z_3) (z_3-z_1)}\,,
\end{equation*}
where $f^{\sa\mathsf{b}\mathsf{c}}$ are the structure constants of $\mathfrak{g}$. There is a similar contribution from the $\tilde{\mathfrak{g}}$ piece.

After evaluating the ghost and worldsheet current algebra portions of the path integral \eqref{bcor2}, it reduces to:
\begin{multline}\label{bcor3}
 f^{\sa_1 \sa_2 \sa_3}f^{\sta_1 \sta_2 \sta_3} \int \frac{[\D\cF]}{\mathrm{vol}\,\C^{*}}\,\e^{-S}\,\left(b^{(2)}|\omega_{0}\right)\,\delta(X^{2}(z_0))\,\prod_{j=1}^{3}\delta\!\left(\log\mathrm{Hol}_{z_j} X^{2}\right) \\
 \prod_{k=1}^{2}\delta\!\left(\mathrm{Res}_{z=z_k} X\cdot P\right)\,\prod_{i=1}^{3} \frac{1}{(k_i\cdot X(z_i))^2}\,,
\end{multline}
where we abuse notation by denoting the remaining path integral measure by $[\D\cF]$. At this point, it is useful to repackage the delta functions in the second line of \eqref{bcor3} as
\begin{equation*}
 \prod_{k=1}^{2}\delta\!\left(\mathrm{Res}_{z=z_k} X\cdot P\right)= \int_{-\infty}^{+\infty} \frac{\d m_{1} \,\d m_{2}}{4\,\pi^2}\,\exp\!\left(i\sum_{k=1,2} m_{k}\,\mathrm{Res}_{z=z_k} X\cdot P\right)\,.
\end{equation*}
The advantage of this representation is that the contribution from these delta functions, now exponentiated, can be taken into the action in \eqref{bcor3}:
\be\label{actmod}
-\frac{1}{2\pi}\int_{\Sigma} P_{\mu}\,\dbar X^{\mu} \;\rightarrow\; -\frac{1}{2\pi}\int_{\Sigma} P_{\mu}\,\dbar X^{\mu} -2\pi\im \sum_{k=1,2} m_{k}\,\mathrm{Res}_{z=z_k} X\cdot P\,.
\ee
With this step, \emph{all} dependence on the field $P_{\mu}$ is isolated in the exponentiated action, and the path integral over $[\D P]$ can be performed explicitly.

Since $P_{\mu}$ has conformal weight $(1,0)$, it has no zero modes, but the path integral over its non-zero-modes imposes an equation of motion on $X^{\mu}$:
\be\label{xeom}
\dbar X^{\mu}(z)= \sum_{k=1,2} m_{k}\,(z-z_{k})\,\dbar\left(\frac{1}{(z-z_k)}\right)\, X^{\mu}(z)\,.
\ee
The right-hand side of this equation is zero as a distribution if $X^{\mu}$ is a smooth function, but it can be non-zero if $X^{\mu}$ has poles or branch cuts. In particular, \eqref{xeom} has a non-trivial formal solution given by
\be\label{xeoms}
X^{\mu}(z)= X^{\mu}_{0}\,\exp\!\left(\sum_{k=1,2}m_{k} \ln (z-z_k)^{-1}\right) = \frac{X^{\mu}_0}{(z-z_1)^{m_1} (z-z_2)^{m_2}}\,,
\ee
where $X^{\mu}_{0}$ is the constant zero mode of $X^{\mu}$. This solution appears to have highly singular behaviour near $z=z_{1},z_{2}$ when $m_{1},m_{2}>0$, but in the context of the path integral the solution must be treated as distributional until the integrals over $\d m_1$ and $\d m_2$ have been performed. So when substituting \eqref{xeoms} back into the path integral, its formality can be dealt with by `regulating' the solution near marked points by the simple prescription
\begin{equation*}
 X^{\mu}(z_i)\rightarrow X^{\mu}(z_i +\varepsilon)\,,
\end{equation*}
for some complex parameter $\varepsilon$ to be taken to zero after all integrals have been performed. 

With this prescription, the three-point correlator is:
\begin{multline}\label{bcor4}
 f^{\sa_1 \sa_2 \sa_3}f^{\sta_1 \sta_2 \sta_3} \lim_{\varepsilon\rightarrow 0} \int \frac{\d^{8}X_{0}\,[\D b^{(2)}]}{\mathrm{vol}\,\C^{*}}\,\int_{-\infty}^{+\infty}\frac{\d m_{1}\,\d m_2}{4\,\pi^2}\,\left(b^{(2)}|\omega_{0}\right)\,\delta(X^{2}(z_0))\\
 \prod_{j=1}^{3}\delta\!\left(\log\mathrm{Hol}_{z_j} X^{2}\right)\,\prod_{i=1}^{3} \frac{1}{(k_i\cdot X(z_i+\varepsilon))^2}\,, 
\end{multline}
where $X^{\mu}$ is given by \eqref{xeoms}. The remaining delta function insertions constrain the holonomy of $X^2$ around each of the marked points; it is easy to see that
\be\label{xhol1}
\mathrm{Hol}_{z_1} X^{2} = \e^{-4\pi\im m_1}\,, \qquad \mathrm{Hol}_{z_2} X^{2} = \e^{-4\pi\im m_{2}}\,,
\ee
while there is no holonomy around $z_3$. This reveals why we must take the \emph{logarithm} of the holonomies: if the holonomy appeared in the delta functions then this would only constraint $m_{1},m_{2}$ to be integer or half-integer. The integrals over $m_1,m_2$ can now be done against two of these delta functions:
\be\label{xhol2}
\int_{-\infty}^{+\infty}\frac{\d m_{1}\,\d m_2}{4\,\pi^2}\,\prod_{j=2,3}\delta\!\left(\log\mathrm{Hol}_{z_j} X^{2}\right)= \int_{-\infty}^{+\infty}\frac{\d m_{1}\,\d m_2}{64\,\pi^4}\,\delta(m_1)\,\delta(m_2)\,,
\ee
thereby setting $X^{\mu}$ equal to its zero-mode everywhere else in the path integral. The third holonomy delta function, $\delta(\log\mathrm{Hol}_{z_3}X^2)$ is singular, but this infinity is cancelled out by the path integral over $b^{(2)}$, which can be used to remove this delta function altogether.

This leaves
\be\label{bcor5}
f^{\sa_1 \sa_2 \sa_3}f^{\sta_1 \sta_2 \sta_3} \int \frac{\d^{8}X_{0}}{\mathrm{vol}\,\C^{*}}\,\frac{\delta(X^{2}_{0})}{64\,\pi^4}\,\prod_{i=1}^{3} \frac{1}{(k_i\cdot X_{0})^2}\,,
\ee
since the $\varepsilon\rightarrow 0$ limit can be taken trivially after all $z_i$-dependence has been removed. The remaining integral is over the projective null cone in $D=8$ dimensions, which is well-defined because the integrand is homogeneous of degree zero.\footnote{This is a consequence of performing the calculation in the critical dimension required to kill the $\SL(2)$ anomalies; if $D\neq 8$ then the resulting integral would not be well-defined.} Finally, the projective integral can be evaluated to give
\be\label{bcor6}
\left\la \prod_{i=1}^{3} V^{[0]}(z_i)\right\ra\ =\frac{1}{64\pi^4} \frac{f^{\sa_1 \sa_2 \sa_3}f^{\sta_1 \sta_2 \sta_3}}{(k_{1}\cdot k_2)\, (k_2\cdot k_3)\, (k_3\cdot k_1)}\,,
\ee
which is precisely the form of the three-point function of single scalar insertions in cubic bi-adjoint scalar theory in six dimensions as required by space-time conformal invariance. It is easy to see that this prescription also gives the correct three-point functions for conformal descendants when vertex operators $V^{[p]}$ for $p>0$ are inserted.

\medskip

In the heterotic model, a similar prescription goes through with some modifications. The correlator with three insertions of \eqref{hVO1} vanishes because the path integral over $\gamma^{(1)}$ zero modes is over constrained (i.e., there are three $\delta(\gamma^{(1)})$ insertions, but only two degrees of freedom in the $\gamma^{(1)}$ zero modes). In string theory, one would remedy this situation by replacing one of the insertions with an operator obtained via the descent procedure from fixed vertex operators. This requires the fixed vertex operator to be proportional to delta functions in all of the ghost fields, or else the resulting descended operator will not generically be BRST-closed. Unfortunately, the vertex operators \eqref{hVO1} do not obey this condition, as they are not proportional to $c^{(2)}$, $c^{(3)}$, or $\delta(\gamma^{(2)})$. Indeed, if one applies the naive descent procedure to \eqref{hVO1}, the result is an operator which fails to be BRST-closed and conflicts with the interpretation of $A_{\mu}$ as a $d=4$ gauge field.

Without a systematic way to obtain the appropriate `descended' vertex operator, we simply make a reasonable ansatz. Consider an operator of the form:
\be\label{dhVO}
U=c^{(0)}\,c^{(1)}\,j^{\sa}\,\psi^{\mu}\,\psi^{\nu}\,F_{\mu\nu}(X)\,,
\ee
where $F_{\mu\nu}$ is anti-symmetric and depends only on $X$. This operator is not obtained by the descent procedure, but has balanced conformal weight and is not proportional to $\delta(\gamma^{(1)})$, so inserting it into the path integral along with two of the vertex operators \eqref{hVO1} no longer over-saturates the ghost zero modes. The operator $U$ is not quite $Q$-closed, but with the constraints $\Box F_{\mu\nu}=0$, $X^{\mu}F_{\mu\nu}=0$, $\partial_{[\sigma}F_{\mu\nu]}=0$, $\partial^{\mu}F_{\mu\nu}=0$, and $X\cdot\partial F_{\mu\nu}=-2 F_{\mu\nu}$, this failure of BRST-closure is reduced to two anomaly terms (both proportional to $\gamma^{(1)}$) which are not in conflict with any of the constraints. These constraints are precisely the conditions for $F_{\mu\nu}$ to represent a $\Delta=2$ field strength in $d=4$ via the $D=6$ embedding space. The remaining lack of $Q$-closure for $U$ should be reflected in any resulting correlation function as a lack of gauge invariance.


So the heterotic analogue of \eqref{bcor1} is the three-point correlator:
\be\label{hcor1}
\left\la V(z_1)\,V(z_2)\,U(z_3)\right\ra\,,
\ee
with the vertex operators given by \eqref{hVO1} and \eqref{dhVO}. Using the prescription:
\begin{multline}\label{hcor2}
\int[\D\cF]\,\e^{-S}\,c^{(3)}(z_{0})\,\left(b^{(2)}|\omega_{0}\right)\,\delta(X^{2}(z_0))\,\prod_{j=1}^{3}\delta\!\left(\log\mathrm{Hol}_{z_j} X^{2}\right) \\
 \prod_{k=1}^{2}\delta\!\left(\mathrm{Res}_{z=z_k} X\cdot P\right)\,V(z_1)\,V(z_2)\,U(z_3)\,,
\end{multline}
it is a fairly straightforward exercise to evaluate the path integral using the same techniques applied in the bosonic model. The result is
\be\label{hcor3}
\left\la V(z_1)\,V(z_2)\,U(z_3)\right\ra\ = f^{\sa_1 \sa_2 \sa_3}\,\int \frac{\d^{6}X_{0}}{\mathrm{vol}\,\C^{*}}\,\delta(X^{2}_{0})\,A_{1\,[\mu}\,A_{2\,\nu]}\,F_{3}^{\mu\nu}\,.
\ee
Since each $A$ is homogeneous of degree $-1$ in $X_0$ and $F$ is homogeneous of degree $-2$ in $X_{0}$, the integral is projectively well-defined only in the critical dimension $D=6$. Furthermore, the result has the correct color and tensor structure for the 3-point function between gauge fields (or their conformal descendants) in $d=4$ Yang-Mills theory, as desired.

The gauge-theory correlator just presented is not gauge invariant, as expected since $U$ is not fully BRST-closed. At linearised level, this can be remedied by taking derivatives to form a correlator of field strengths. A complete correlator of non-linearly gauge-invariant operators would require us to extend our analysis to composite operators and to properly understand the descent procedure for the model, which is beyond the scope of this paper.

\medskip

For the type II model, one can try to implement a similar prescription. However, the answer one obtains is always zero since there is a new fermionic zero mode to contend with, associated with the field $\eta$. Attempting to remedy this by including an explicit insertion of $\eta$ in the path integral (akin to how the $c^{(3)}$ zero-mode is saturated) still gives zero, since this results in an additional quotient by vol $\C^*$ in the answer. 

Although obtaining zero for the three-point function might seem an indication that the general prescription is breaking down, this is consistent with the fact that $d=2$ Einstein gravity is topological. Indeed, the constraints \eqref{2VO2} imposed on $h_{\mu\nu}$ in $D=4$ are actually satisfied by \emph{any} $d=2$ traceless, symmetric tensor. It would be important to study more closely the spectrum of our type II model to understand if it possesses other states.


\section{Discussion}
\label{Disc}

Let us summarize our findings. We proposed a description of certain $d$-dimensional classical CFTs in terms of models governing holomorphic maps from the Riemann sphere to the phase space of the projective null cone inside $(d+2)$-dimensional Minkowski space. The restriction to the target space is enforced by bosonic constraints, which form a $\SL(2)$ algebra. By quantizing the model on the Riemann sphere, we found that the anomaly from this $\SL(2)$ algebra of constraints vanishes if and only if the space-time dimensionality $d$ is consistent with conformal symmetry, given the spin of the space-time fields naturally associated to the vertex operators of the model. In particular, we found a `bosonic' model for a (biadjoint) cubic scalar field in $d=6$, a `heterotic' model for a Yang-Mills field in $d=4$, and a `type II' model for Einstein gravity in $d=2$. These models differ in the worldsheet matter content and in supersymmetric extensions of the algebra of constraints. For each one, we identified in the spectrum of vertex operators the full conformal multiplet of single field insertions. We showed how the ambitwistor strings describing the same space-time theories are obtained by a particular gauge fixing, whereby the algebra of constraints is reduced. Finally, we gave a prescription for the computation of three-point functions in these models.

Our work leaves many questions unanswered. The first priority is to show how to use our models to compute general (tree-level) correlation functions, beyond three-point examples. In order to do this, we need a more careful and systematic understanding of the gauge-fixing procedure. It is also likely that beyond three points, we will need a different way of performing the path integral over fields, as suggested by field theory computations \cite{mfpToappear}.
The natural expectation is that the models will lead to expressions for CFT correlation functions analogous to the CHY formulae for scattering amplitudes, which can be obtained from ambitwistor strings. In fact, recent work has led to CHY-type formulae for form factors~\cite{He:2016dol,Brandhuber:2016xue,He:2016jdg}, and the formalism that comes from of our models should be somewhat related. 

It is also natural to wonder whether quantum effects can also be described by our models. We argued that certain anomalies in each of the theories prevent their formulation on higher-genus Riemann surfaces, which are the natural setting for loop corrections in a string theory. This could be related to the fact that the spacetime CFTs we obtain do not preserve conformality at loop level. Nonetheless, we also pointed out that the appropriate loop expansion in models of this type may be based instead on Riemann spheres with nodes (pairs of identified points; one pair per loop order), as suggested by the results of \cite{Geyer:2015bja,Geyer:2015jch,Geyer:2016wjx} for ambitwistor strings. Further investigation is required to test this possibility, both in our models and in ambitwistor strings.

One (potentially related) question regards the spectra of the models and their comparison with the space-time CFT. We were able to identify only the conformal multiplets of single field insertions from the space-time CFTs in our models; it remains an important task to account for the multitude of composite operators with their spectrum of conformal dimensions. While it is true that these can always be accessed by taking OPE limits of single field correlators, it would be nice if there would be a more direct way to obtain them. Perhaps the ``unwanted'' vertex operators in our 2d models, which have no clear interpretation in terms of the space-time CFT, could come to the rescue. Note that this question is intimately related to whether disconnected contributions to correlation functions can be obtained from our models, since these can become connected contributions for composite operators. Resolving these issues is a crucial obstacle to interpreting our models as a full, 2d description of space-time CFTs. 

Another possibility suggested by ambitwistor strings (and their predecessors) is the existence of models whose 2d fields are specific to the space-time dimensionality. As shown originally in~\cite{Witten:2003nn}, and more recently for ambitwistor strings in \cite{Geyer:2014fka}, twistor fields naturally describe $d=4$ physics in worldsheet models, and have been particularly useful for $\cN=4$ SYM theory. Given the tremendous progress in computing correlation functions in $\cN=4$ SYM, it would be interesting to explore a twistor-like analogue of our `heterotic' Yang-Mills model.

To conclude, our results provide a new example of how space-time physics can be encoded in 2d models. This basic concept is at the heart of both string theory and the more recent worldsheet models for quantum field theories. We expect that this idea will lead to many more surprises.

\section*{Acknowledgments}

We would like to thank David Skinner and Ellis Ye Yuan for interesting conversations. TA is supported by an Imperial College Junior Research Fellowship. RM is supported by a Royal Society University Research Fellowship. MFP acknowledges funding from the European Community's 7th Framework Programme in the form of a Marie Curie Intra-European Fellowship, PIEF-GA-2013-623606.

\appendix
\section{Gauge fields and gravitons from embedding space}
\label{appendix}

To compute objects in $d$-dimensional space one acts with pullbacks on embedding space expressions,
\be
T_{ab\ldots}=\frac{\partial X^\mu}{\partial x^a}\frac{\partial X^\nu}{\partial x^b}\ldots T_{\mu \nu \ldots} \bigg|_{X^2=0} \,.
\ee
We should demand that each $D$-dimensional index is transverse, i.e. $X^\mu T_{\mu \ldots}=0$. This, together with the invariance of the expression above under shifts of $T_{\mu \nu \ldots}$ by $X_\mu \lambda_{\nu \ldots}$, guarantees that there are only $d-2$ physical components for each index. In order to impose transversality, one can apply the projector:
\be
U^{\mu \nu}=\eta^{\mu \nu}-\frac{X^\mu I^\nu}{X\cdot I}-\frac{X^\nu I^\mu}{X\cdot I}+X^2 \frac{I^\mu I^\nu}{(X\cdot I)^2}\,,
\ee
which satisfies $U^{\mu \nu} X_\nu=0$, and incidentally is the same as the $d$-dimensional metric, as can be seen by acting with the pullbacks. The explicit presence of the $I$ vectors even after imposing $X^2=0, X\cdot I=1$, implies that this is not a conformally invariant object.

As a warm-up, let us compute the $d$-dimensional Laplacian in embedding space. We write
\be
\partial_a \partial^a \phi(x)=\eta^{\mu \nu}U_{\mu}^{\ \rho} U_{\nu}^{\ \tau}\partial_\rho\left( U_{\tau}^{\ \gamma}\partial_\gamma \Phi(X)\right) \bigg|_{X^2=0}\,.
\ee
We can safely set $X\cdot I=1$ and $X^2=0$ everywhere, thanks to the presence of the projectors. The computation gives
\begin{eqnarray}
\eta^{\mu \rho}U_{\mu}^{\ \nu} U_{\rho}^{\ \tau}\partial_\nu\left( U_{\tau}^{\ \gamma}\partial_\gamma \Phi(X)\right)
&=&-d\, I\cdot \partial \Phi+ \partial_\mu\partial^\mu \Phi-2 I^\mu X^\nu \partial_\mu \partial_\nu \Phi \nonumber \\
&=& \partial_\mu\partial^\mu \Phi-(d-2\Delta-2) I\cdot \partial \Phi\,,
\end{eqnarray}
where we assumed $X\cdot\partial_X \Phi=-\Delta \Phi$. Notice that for $\Delta=(d-2)/2$ the dependence on the $I$ vector drops out and the flat space expressions simply lift to embedding space directly. This is the case since for this value of $\Delta$ we have that $\phi$ is a free, conformally invariant field, and in particular its descendant $\Box \phi$ is also a 
(null) primary. The same computation gives
\be
\partial_a V^a(x)= \partial_\mu V^\mu-(d-1-\Delta) I\cdot V \,,
\ee
and again the dependence on $I$ drops out when $\Delta=d-1$, i.e. $V^a$ is a conserved current, whose divergence is again a (null) conformal primary.

\medskip

Let us now see what kind of field is determined by the conditions (\ref{hVO2}). One computes:
\begin{eqnarray}
\partial_c \partial^c A_a(x)&=&\frac{\partial X^{\mu}}{\partial x^a} \left[ \partial_\nu \partial^\nu A_\mu-2 \partial_\mu I\cdot A+(2\Delta+2-d) I\cdot \partial A_\mu\right] \,, \\
\partial_c \partial_a A^c(x) &=&\frac{\partial X^{\mu}}{\partial x^a} \left[ \partial_\nu \partial_\mu A^\nu-(d-1-\Delta) I\cdot A\right] \,.
\end{eqnarray}
This implies
\begin{eqnarray}
\partial_c F^{c}_{\ a}=\frac{\partial X^{\mu}}{\partial x^a}\left[\partial_\nu F^{\nu}_{\ \mu}+(d-3-\Delta)\partial_\mu I\cdot A+(2\Delta+2-d) I\cdot \partial A_\mu \right]\,,
\end{eqnarray}
where we have defined field strengths, e.g. $F_{ab}=\partial_a A_b-\partial_b A_a$. For $\Delta=1, d=4$, the dependence on $I$ completely drops out, as it should, since in this case $\partial_a F^{ab}$ behaves as a conserved current. Altogether, the conditions (\ref{hVO2}) imply that $A_a(x)$ satisfies the free gauge field equation of motion in $d=4$.

\medskip

Now let us turn to spin two. One computes, denoting $\xi^{\mu}_a=\partial X^\mu / \partial x^a$,
\begin{eqnarray}
\partial_a \partial^a h_{bc}&=&\xi^{\mu}_b \xi^{\nu}_c \left[ \partial^2 h_{\mu\nu}-4 I^\rho \partial_{(\mu} h_{\nu) \rho}+(2\Delta+2-d)\,I\cdot \partial h_{\mu\nu}+2\eta_{\mu\nu} I^{\rho}I^{\tau}h_{\rho\tau}\right] \,, \nonumber \\
\partial_a\partial_b h^{ab}&=&\partial_\mu\partial_\nu h^{\mu\nu}+(d-1-\Delta)\left[-2 I^\rho \partial_\tau h^{\tau}_{\ \rho}+(d-\Delta) I^{\alpha}I^{\beta}h_{\alpha \beta}\right]+I\cdot \partial h \,, \nonumber \\
\partial_a \partial_c h^{c}_{\ b}&=&\xi_{a}^\mu \xi_{b}^{\nu}\bigg(\partial_\mu\partial_\rho h^{\rho}_{\ \nu}-(d-\Delta)\left(I^\rho\partial_\mu h_{\rho\nu}+\eta_{\mu\nu}I^\rho I^\tau h_{\rho \tau}\right)-\eta_{\mu\nu}I^\rho\partial^\tau h_{\rho\tau}\bigg) \,.
\end{eqnarray}
Setting the $D$-dimensional trace to zero, one obtains that the following quantity is conformally invariant, i.e. it can be lifted directly to embedding space, only for $d=2$ and $\Delta=0$:
\be
R_{ab}:=\Box h_{ab}-2 \partial_c \partial_{(a} h_{b)}^{\ c}+\eta_{ab} \partial_{c}\partial_d h^{cd} \,.
\ee
This is of course nothing but the linearized Einstein tensor in two dimensions. The conditions (\ref{2VO2}) then imply that $R_{ab}=0$, with the further condition that $h_{ab}$ should be traceless.

\bibliography{notes}
\bibliographystyle{JHEP}

\end{document}